\documentclass[preprintnumbers,
amsmath,amssymb,floatfix,10pt,prd,onecolumn,
superscriptaddress,longbibliography,nofootinbib]{revtex4-2}
\usepackage{bm}
\usepackage{amsfonts}
\usepackage{latexsym}
\usepackage{amsmath}
\usepackage{textcomp}
\usepackage{float}
\usepackage{booktabs}
\usepackage{dcolumn}
\usepackage{dsfont}
\usepackage{ragged2e}
\usepackage{epsfig}
\usepackage[dvipsnames]{xcolor}
\usepackage{hyperref}
\hypersetup{           
	colorlinks=true,                
	breaklinks=true,                
	urlcolor= OrangeRed,                
	linkcolor= magenta,                
	bookmarksopen=false,
	filecolor=black,
	citecolor=red,
	linkbordercolor=blue}
\usepackage{graphicx}
\usepackage{orcidlink}
\usepackage[T1]{fontenc}

\begin{document}
	\title{Modeling a Non-Singular Universe with Late-Time Acceleration through a Novel Inhomogeneous Barotropic Equation of State}
	
	\author{Rajdeep Mazumdar \orcidlink{0009-0003-7732-875X}}%
	\email{rajdeepmazumdar377@gmail.com}
	\affiliation{%
		Department of Physics, Dibrugarh University, Dibrugarh \\
		Assam, India, 786004}

	\author{Mrinnoy M. Gohain\orcidlink{0000-0002-1097-2124}}
	\email{mrinmoygohain19@gmail.com}
	\affiliation{%
		Department of Physics, Dibrugarh University, Dibrugarh \\
		Assam, India, 786004}
	
	\author{Kalyan Bhuyan\orcidlink{0000-0002-8896-7691}}%
	\email{kalyanbhuyan@dibru.ac.in}
	\affiliation{%
		Department of Physics, Dibrugarh University, Dibrugarh \\
		Assam, India, 786004}%
	\affiliation{Theoretical Physics Divison, Centre for Atmospheric Studies, Dibrugarh University, Dibrugarh, Assam, India 786004}

	
	\begin{abstract}
	In this study, we investigated the effects of incorporating barotropic fluids on cosmological solutions within the general relativity (GR) framework. We proposed a modified version of the barotropic fluid with the EoS, $p=\zeta _0 \rho +\zeta _1 \rho \left(t-t_0\right){}^{-2 n}$, where $\zeta_0$, $\zeta_1$, $t_0$ and $n$ are some constants. Our goal is to explore if this type of EoS might help explain the universe's development, concentrating on the scenario where the universe bounces instead of singularities. Interestingly the generic solutions derived from our model are sufficiently adaptable to illustrate the bounce scenario, cosmic inflation and late-time dark-energy behaviour. The parameters $\zeta_0$, $\zeta_1$, $t_0$, and $n$ define the universe's phase in this non-singular solution. We investigated several elements of cosmic development, including as the energy density, deceleration parameter, and energy conditions, in order to validate our model. Stability analysis showed that the perturbations approach to zero as the time evolves, indicating the model is stable under scalar perturbation. Additionally, we looked at the statefinder diagnostics and Hubble flow dynamics to get more understanding of the model's dark energy and inflationary behaviour, respectively. Additionally, we conducted a study of the models' relevance to the observational datasets from BAO, DESI and Pantheon+SH0ES.
	\end{abstract}
	
	\maketitle

\section{Introduction}\label{sec1}
Our knowledge of spacetime and gravity was completely transformed by Einstein's general relativity (GR), which is essential to modern era of physics. GR is subsequently the essential foundation to the astrophysical and cosmological world \cite{ref1}. Phenomena like gravitational waves, black holes, quasars, and even the cosmic dynamics of the entire universe might all be physically explained by GR, a geometric theory that is mathematically simple and has a lot of potential \cite{ref1, ref2}.  A number of works have been developed and re-created based on GR (see Ref. \cite{ref3} to Ref. \cite{ref29}). The fundamental idea of a common cosmological perspective is that the universe began with the Big Bang, defined as an initial space-time singularity around a point of infinite energy density that purely requires knowledge of quantum gravity. However, apart from the issue of initial singularity , conventional cosmology based on Big Bang presents several other issues, such as the trans-planckian problem, the horizon problem, and the flatness problem \cite{ref30,ref31}. Here, Guth's inflation hypothesis provided effective explanation in primly resolving the majority of such issues \cite{ref32}. However, the trans-Planckian problem of fluctuations and the singularity problem afflict the inflationary scenario \cite{ref31}. Since a strong and scientifically valid theory of quantum gravity is still absent, the inflationary scenario fails as a comprehensive cosmological theory at the beginning of inflation, when the universe grows nearly exponentially. To overcome the aforementioned issue, scenarios alternative to the inflation like pre-big bang \cite{ref33}, emergent scenario \cite{ref34,ref35,ref35_a}, ekpyrotic/cyclic scenarios \cite{ref36,ref37}, and last but not least bounce scenario \cite{ref38,ref38_a} were proposed.\\
The family of theories known as "bouncing cosmologies" \cite{ref39} enables the study of the universe at or prior to the shift to conventional cosmological history by removing the necessity for a theory of quantum gravity.  Instead than retracing the expansion back to an infinitesimal point, bouncing cosmologies assume that the universe existed at a non-minimal size and restricted energy density at one point, beyond which contraction was impossible. As a result, the universe avoids the singularity by emerging from the expansion of the previous contracting phase rather than from a singularity. In the idea of bouncing cosmology, the Hubble parameter increases sharply after hitting zero, but the scale factor drops to a certain extent before expanding. The primary benefit of bouncing cosmology is its ability to resolve the mainstream cosmological paradigm's singularity issue by substituting a smooth contraction-to-expansion transition (Big Bounce) for the singularity (Big Bang). This makes it possible to see the early cosmos in a more continuous manner. The wedge diagram put out by Ijjas and Steinhardt \cite{ref40} provides a visual representation of the efficacy and efficiency of bouncing models in addressing fundamental cosmological issues. The null energy condition must theoretically be broken in order for a bounce scenario to be realised. The Hubble parameter $H$ grows with time, and $\dot{H} > 0$, according to the NEC violation. We go into further detail below on the necessary characteristics or requirements of bouncing cosmic scenarios \cite{ref41}:
\begin{itemize}
    \item During the bouncing epoch, the deceleration parameter subsequently becomes singular, the Hubble parameter fades to zero and the scale factor compresses to some non-zero finite value.
    \item The NEC gets violated, since the a shift in the sign of Hubble parameter occurs at the bouncing point; consequently, this phenomena is excluded from consideration in the framework of General Relativity (GR).
    \item The Hubble parameter stays negative during contraction era and positive during expansion , although the slope of the scale factor rises following the bounce.
\end{itemize} 
One can go through Ref. \cite{ref42,ref43,ref44,ref45} (and references within) and in Ref. \cite{ref46,ref46a,ref46b,ref46c,ref46d,ref46e,ref46f} for a detailed review of numerous studies that have been conducted in the literature using various frameworks, such as scalar fields, quintom matter, $f(T)$ gravity, and string inspired gravity. Up till now, several bouncing cosmologies have been developed, including matter bounce, oscillatory, super bounce, and symmetric bounce. Here, the symmetric bounce had been initially adopted to create a non-singular bouncing cosmology after an ekpyrotic phase of contraction by Cai et al. \cite{ref44}. Primordial modes in this bounce, however, have problems since they don't reach the Hubble horizon unless they are combined with other cosmic behaviours. \cite{ref47,ref48,ref49}. In order to formulate an universe without a singularity and in which the universe collapses and raises through a Big Bang, Koehn et al. \cite{ref50} first proposed super bounce cosmologies \cite{ref51}. Such type of cosmology is illustrated by the power-law type scale factor. It should be noted that the super bounce is also associated with a Hubble parameter that changes signatures in eras both before and after the instance of bounce, but retains a vanishing value at the bounce point. Oscillating bouncing cosmologies are given by an oscillating scale factor. A cyclic universe, which views the world as a constant cycle of contractions and expansions, is reflected in the behaviour of such models \cite{ref52,ref53,ref54,ref55,ref56}. The matter bounce cosmology formulated based on the idea of the loop quantum cosmology (LQC) has been employed to explore the early universe and it is able to produce a power-spectrum that is scale-invariant or nearly scale-invariant \cite{ref57,ref58,ref59}.\\
The scenario of an accelerating universe has been greatly backed by the insights on observable data\cite{ref60}. Cosmological acceleration can be introduced via exotic fluid having negative pressure (dark energy) \cite{ref61, ref62}  or via modification of gravity \cite{ref63}. And, among all the alternate models and theories barotropic fluid is also a widely studied one. The barotropic fluid, in general terms can be some form of field or matter, with the energy density $\rho$ and pressure $p$ which satisfies the relation $p = f(\rho)$. Here, $f(\rho)$ represents some arbitrary or function, and we refer to this relationship as what we popularly call the equation of state (EoS). A variety of evolutionary features of the visible universe have been described using barotropic fluids. Some specific and viable instances of barotropic fluids in literature are like the affine EoS \cite{ref64,ref65}, quadratic EoS \cite{ref66,ref67}, Chaplygin gas along with its modified forms \cite{ref68,ref69}, logotropic EoS \cite{ref70,ref71}, polytropic fluid \cite{ref72,ref73}, and finally the Van der Waals EoS \cite{ref74,ref75}. One can consult Ref. \cite{ref75a, ref75b, ref75c,ref75d,ref75e} for more prior associated works. In some cases an equation of state which are phenomenological in nature may also be employed, that comprises of The Hubble parameter (\( H \)) and its derivatives, as well as the energy density, given by  $p = f(\rho, H, \dot{H}, \ddot{H}, \dots)$. These equations of state, which are included in the most general models of dark fluids, can be referred to as the inhomogeneous EoS \cite{ref76,ref77}. The bulk viscosity that have time dependence  may also give rise to the inhomogeneous equations of state, which might also push the cosmos into the phantom era \cite{ref77,ref78}. There has also been discussion of an inhomogeneous time-dependent EoS for the dark energy as the fluid-like description that corresponds to modified theories of gravity. (see Ref. \cite{ref79}). In this paper, we investigate how the introduction of a modified EoS impacts the cosmic development of the universe. Here, we focus on the possibility of obtaining a general non-singular cosmological solution with late-time acceleration, which is physically viable and consistent with the observed cosmological data respectively.\\
The following is how the work is presented. In Sec. \ref{sec2}, we obtain the general solutions to the Friedmann equations for a model with a modified EoS $p=\zeta _0 \rho +\zeta _1 \rho  \left(t-t_0\right){}^{-2 n}$. The evolution of physical cosmology observables such as the deceleration parameter and energy density for the model, as well as the relevant energy conditions to verify the model, are examined in Sec. \ref{sec3}. We examine the model's stability in Sec. \ref{sec4.1}. In Sec. \ref{sec4} and Sec. \ref{sec5}, we study the Hubble flow dynamics and statefinder diagnostics for the model respectively. We examine how well the model is able to account for the observational data in Sec. \ref{sec6}. In Sec. \ref{sec7} we conclude the study with our findings and future perspectives. 

\section{Framework and Solution}\label{sec2}
\begin{figure*}[htb]
\centerline{\includegraphics[scale=0.55]{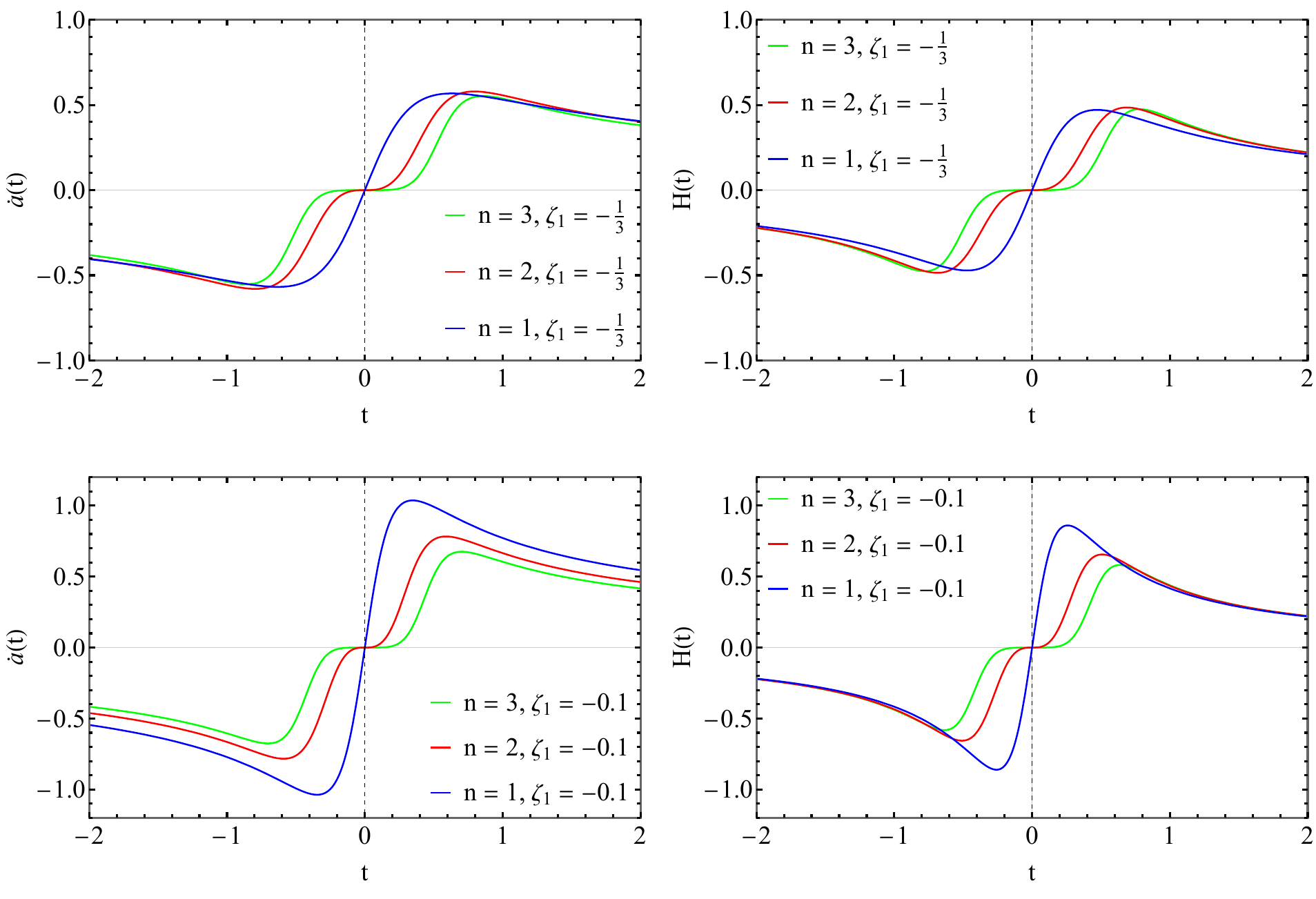}}
\caption{Plot shows evolution of the Hubble parameter and Scale factor's first-order time derivative with cosmic time for the model for various values of the parameter \(n\). We have made the following assumptions for the plot: \(t_0 = 0\), \(a_0 = 1\), and \(zeta_0 = 0.5\) for the plot. $t$ and $\dot{a}$ have the units $Gyr$ and $Gyr^{-1}$, respectively.}
\label{fig1}
\end{figure*}

\begin{figure*}[htb]
\centerline{\includegraphics[scale=0.55]{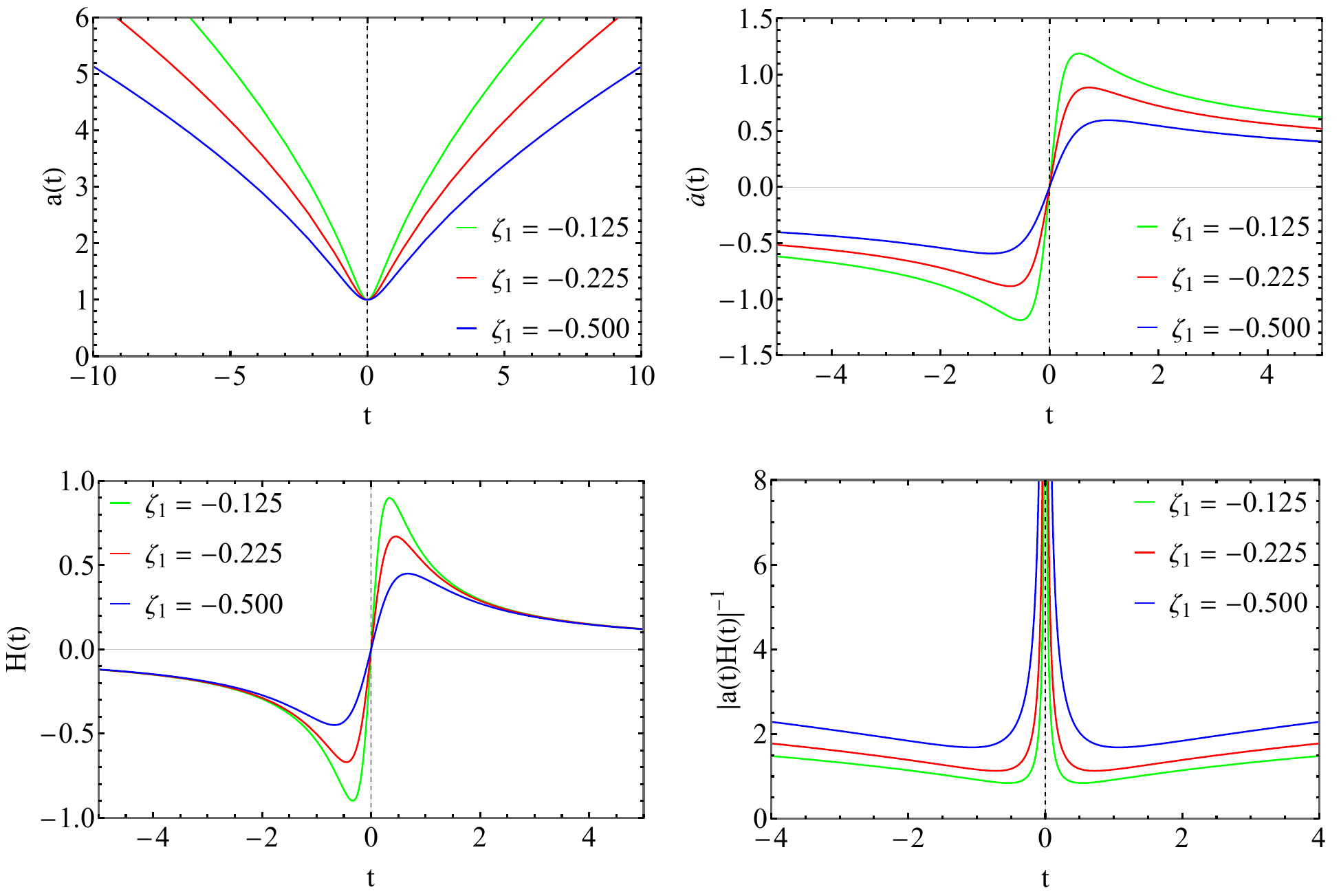}}
\caption{Evolution with cosmic time in the scale factor, Hubble parameter, Hubble radius, and the scale factor's first-order time derivative. We have made the following assumptions for the plot: \(\zeta_0=0.5\), \(n=1\), \(t_0=0\), and \(a_0=1\). $Gyr$ and $Gyr^{-1}$ are the units of $t$ and $H$, respectively.}
\label{fig2}
\end{figure*}

\begin{figure*}[htb]
\centerline{\includegraphics[scale=0.55]{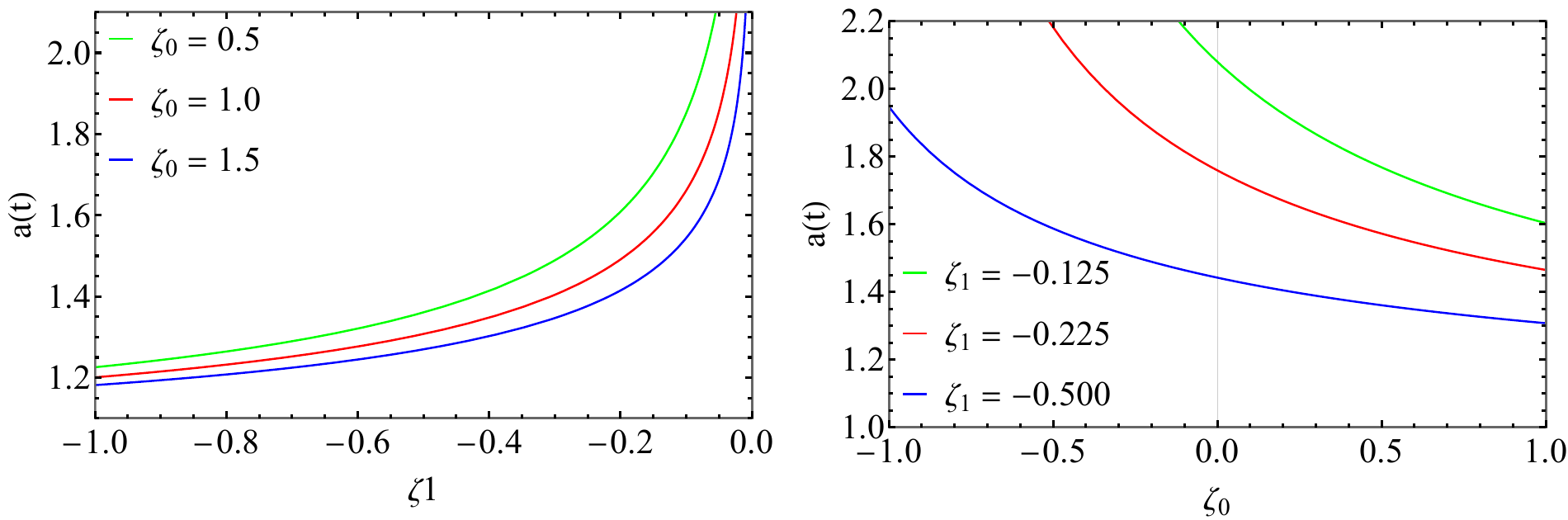}}
\caption{Plot of variation of the Scale factor with respect to different values of the model parameter \(\omega\) and \(\zeta_0\). We have made the following assumptions for the plot: \(t = 1\), \(n = 1\), \(t_0 = 0\), and \(a_0 = 1\).}
\label{fig3}
\end{figure*}

\begin{figure*}[htb]
\centerline{\includegraphics[scale=0.55]{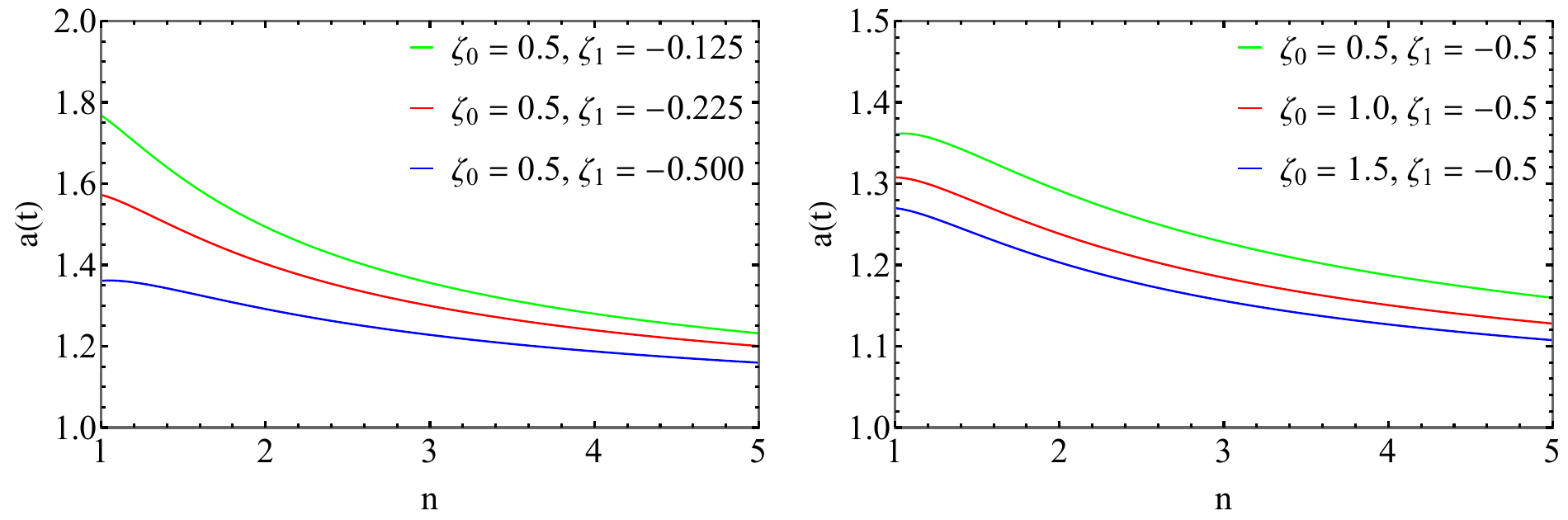}}
\caption{Plot of variation of the Scale factor with respect to different values of the model parameter \(n\). We have made the following assumptions for the plot: \(t = 1\), \(t_0 = 0\), and \(a_0 = 1\).}
\label{fig4}
\end{figure*}

The field equations in GR, which link space-time geometry to the energy and matter content is given by:
\begin{equation}
    G_{\mu\nu} = 8 \pi G T_{\mu\nu}. \label{eq:1}
\end{equation}
Here, the curvature of space-time is represented by the Einstein curvature tensor, $G_{\mu\nu}$; the distribution of matter and energy in space-time is represented by the energy-momentum tensor, $T_{\mu\nu}$; the geometry of space-time is described by the metric tensor, $g_{\mu\nu}$; and the gravitational constant, $G$, is the gravitational constant, respectively.  The FLRW line element, which is spatially flat, homogeneous, and isotropic, is used here by:
\begin{equation}
    ds^2 = -dt^2 + a(t)^2(dx^2 + dy^2 + dz^2). \label{eq:2}
\end{equation}
Here, \(a(t)\) is the scale factor. Hence, from Eq. (\ref{eq:1}) and Eq. (\ref{eq:2}), the field equations that explains the evolution of the scale factor \(a(t)\) are governed by the following equations:
\begin{equation}
    H^{2} = \left(\frac{\dot{a}}{a}\right)^2 = \frac{\rho}{3}, \label{eq:4}
\end{equation}
\begin{equation}
    \frac{\ddot{a}}{a} = \dot{H} + H^2 = -\frac{1}{6}(\rho + 3 p).\label{eq:5}
\end{equation}
In the energy-momentum tensor $T_{\mu\nu} = (\rho + p)u_\mu u_\nu + p g_{\mu\nu}$, where $u^\mu = (-1,0,0,0)$ is the fluid's four-velocity that meets the condition $u^\mu u_\mu = -1$, $H$ stands for the Hubble parameter, and $\rho$ and $p$ represent the energy density and pressure, respectively.  We have assumed $8\pi G =1$, where the derivative with respect to time is indicated by the overhead dot (.).  These equations describe how the expansion of the universe is impacted by the cosmological constant, space curvature, and the energy density of matter and radiation.\\
By taking into consideration a modified barotropic EoS, we now concentrate on the potential for obtaining an accurate solution to the field equations provided by Eqs. (\ref{eq:4}) and Eq. (\ref{eq:5})  in the presence of a barotropic fluid. In different works, EoS of different forms have been taken under consideration to explain the evolution of the universe and it's corresponding features (see Ref.\cite{ref64} to Ref. \cite{ref78} for a detail review). Barotropic EoS are often interpreted in cosmological situations as functions of energy density, the Hubble parameter, or higher powers or derivatives of these, which can later be interpreted as functions of time or acceleration- or velocity-dependent variables \cite{ref79}. Under the light of which, let us propose a new parametrization of the  barotropic EoS : 
\begin{equation}
   p=\zeta _0 \rho +\zeta _1 \rho  \left(t-t_0\right){}^{-2 n}, \label{eq:6}
\end{equation}
where, $\zeta_0$, $\zeta_1$, $t_0$ and $n$ are some constant parameters. Here, $n$ and $\zeta_0$ are dimensionless, $t_0$ have the dimension of time and similarly $\zeta_1$ have the dimension $[M]^{1}[L]^{-1}[T]^{2n}$ to validate the dimensional correctness of the Eq. (\ref{eq:6}) throughout the work. Such a form of inhomogeneous time-dependent EoS may be special in a sense that, it can provide a unified picture to explain a non-singular early universe and the late time accelerated expansion \cite{refm}. Then using Eqs. (\ref{eq:4}) and (\ref{eq:6}), we can redefine Eq. (\ref{eq:5}) into the following form:
\begin{equation}
    \dot{H} = (\alpha (t - t_0)^{-2n} - \beta) H^2. \label{eq:7}
\end{equation}
Here, \(\alpha = -\frac{3}{2} \zeta_1\) and \(\beta = \frac{3}{2} (1 + \zeta_0)\). Solving Eq. (\ref{eq:7}), gives the Hubble parameter as:
\begin{widetext}
\begin{equation}
    H(t) = \frac{(-1 + 2n)(t - t_0)^{2n}}{-t_0\alpha + t(\alpha + (-1 + 2n)(t - t_0)^{2n}\beta) - (-1 + 2n)(t - t_0)^{2n}C_1}. \label{eq:8}
\end{equation}
\end{widetext}
\(C_1\) being some constant. Now, we know \(H = \frac{\dot{a}}{a}\), hence integration of Eq. (\ref{eq:8}) can help to obtain the scale factor as:
\begin{equation}
    a(t) = C_2 \left(n(\alpha + (-1 + 2n)(t - t_0)^{2n}\beta)\right)^{\frac{1}{2n\beta}}, \label{eq:9}
\end{equation}
where we make the assumption that \(C_1 = t_0\beta\) to obtain a solution in somewhat close form, with \(C_2\) again being some constant, and subjected to the condition \(a(t) \to a_0\) when \(t \to t_0\), we get \(C_2 = a_0 \left(n\alpha\right)^{-\frac{1}{2n\beta}}\), thus we have:
\begin{equation}
    a(t) = a_0 \left(n\alpha\right)^{-\frac{1}{2n\beta}} \left(n(\alpha + (-1 + 2n)(t - t_0)^{2n}\beta)\right)^{\frac{1}{2n\beta}}. \label{eq:10}
\end{equation}
As \(H = \frac{\dot{a}}{a}\)  we can rewrite the expression for the Hubble parameter as:
\begin{equation}
    H(t) = \frac{(2 n-1) (t-t_0)^{2 n-1}}{\alpha +\beta  (2 n-1) (t-t_0)^{2 n}}. \label{eq:10.1}
\end{equation}
The scale factor rises (\(\dot{a} > 0\)) during the expanding phase (positive time zone) and falls (\(\dot{a} < 0\)) during the contracting phase (negative time zone) across a range of parameter values, as shown in Figs. \ref{fig1} and \ref{fig2}.  At \(t = t_0\), the outcome is \(a \neq 0\) and \(\dot{a} = 0\).  At \(t = t_0\), the Hubble radius is observed to diverge, and the Hubble parameter changes from a negative value (Contracting Universe) to a positive one (Expanding Universe). So, as it satisfies the basic conditions needed for a bouncing cosmology, it seems evident enough that the model and the scale factor derived from it represent a bouncing cosmology in which the bounce occurs at the some cosmological time \(t = t_0\). Here, now we can define the parameter \(t_0\) as the cosmological time at which the bouncing epoch occurs, for the plotting purpose we had taken \(t_0 = 0\).
We have numerically examined the scale factor's variation with the model parameters at cosmological time \(t = 1\) for \(a_0 = 1, t_0 = 0\) as illustrated in Fig. \ref{fig3} and Fig. \ref{fig4} for clearly comprehending the dependence of the scale factor associated to the model on \(\zeta_1\), \(\zeta_0\), and \(n\) the model parameters. From which it follows that, the scale factor rises as the value of the parameter $\zeta_1$ grows and falls as the value of the parameter $\zeta_0$ increases. Furthermore, it is noted that there is a strong influence of the scale factor on the parameter \(n\).\\
Conclusively from the analysis it's is relevant that the model under our consideration is fulfilling the basics needs of a bouncing universe scenario, but it is also important that the model is able to satisfy other necessary conditions for representing a physically viable scenario of non singular cosmology enclosing the idea of bouncing universe along with late time accelerated dark energy behaviour. For which in the following sections we try to evaluate the evolution of cosmological observables and energy conditions for the model, stability analysis along with the Hubble flow dynamics, statefinder diagnostics and comparison with latest observable data.

\section{Evolution of Cosmological observables and Energy Conditions}\label{sec3}
In order to evaluate the validity of the model, in this section we try to look at both the important energy conditions it reflects and the evolution of some cosmological observables like the energy density and the deceleration parameters.
\subsection{Energy Density}
\begin{figure*}[htb]
\centerline{\includegraphics[scale=0.55]{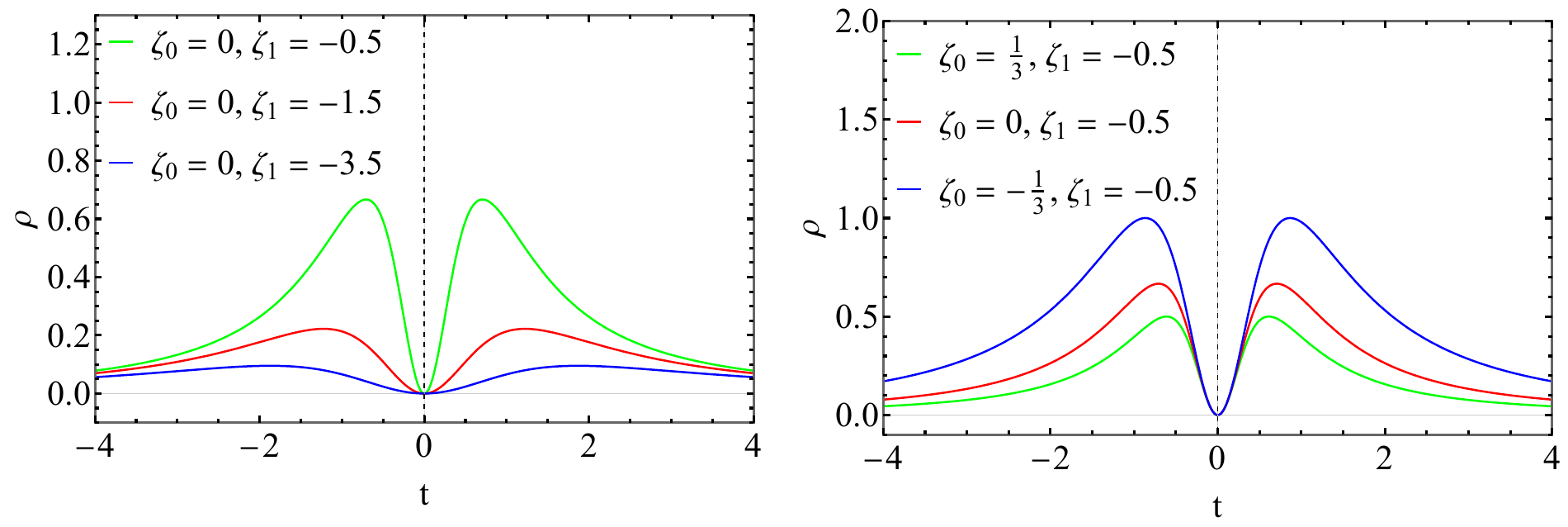}}
\caption{Variation of the energy density with cosmological time \( t \). We have made the following assumptions for the plot: \( n = 1 \), \( t_0 = 0 \), and \( a_0 = 1 \). The unit of $t$ is in $Gyr$.}
\label{fig5}
\end{figure*}

Eqs. (\ref{eq:4}) and (\ref{eq:10}), can help us to calculate the energy density for the model under consideration as:
\begin{equation}
    \rho = \frac{4}{3 (t - t_0)^2 \left(1 + \left(\frac{(t - t_0)^{-2n} \zeta_1}{-1 + 2n} + \zeta_0\right)^2\right)}. \label{eq:11}
\end{equation}
For various combinations of model parameters, Fig. \ref{fig5} illustrates how the energy density \(\rho\) changes with the cosmic time \(t\). Observations show that during the phase of contraction, the energy density first rises as time increases until it reaches some maximum value. Following this, the energy density begins to fall as time increases, eventually reaching zero at the bouncing epoch. 
The energy density for the phase of expansion on the other hand starts at zero during the bouncing epochs, rises with increasing time until it reaches a maximum value once again, and then falls with increasing time.  However, the energy density continuously maintains a non-negative value as required and anticipated.  The loss of the energy density at that point is caused by the violation of the Null Energy Condition (NEC) at the bouncing epoch. As a result, the energy density decrease at the bouncing point as shown in Fig. \ref{fig5} suggesting that the NEC is broken. The results appear to be comparable to those found in other studies \cite{ref80,ref81,ref82,ref83}. A detailed discussion into the NEC violation is given in Sec. \ref{sub3}

\subsection{Deceleration Parameter}
The deceleration parameter in cosmology is a parameter that indicates how rapidly the universe is slowing down its expansion. It gives a measure of the evolution of the universe's expansion rate throughout time. The deceleration parameter \( q \) is defined as:
\begin{equation}
    q = -\frac{\ddot{a}a}{\dot{a}^2}. \label{eq:12}
\end{equation}
A deceleration occurs for \( q > 0 \), and accelerating universe situations occur for \( q < 0 \). Cosmological models are classified into the following groups according to Singh \& Bishi's \cite{ref84} study, as indicated in Table \ref{tab:1}, according to their temporal dependency on the Hubble parameter and deceleration parameter, respectively. According to the categorisation, instances A, B, and C might occur as in the present case universe seems to be expanding. Depending on it, our universe exhibits various forms of expansion, as given in Table \ref{tab:2} respectively \cite{ref84}. Using Eq. (\ref{eq:10}) and Eq. (\ref{eq:12}), we obtain the deceleration parameter \( q \) for the model as:
\begin{equation}
    q = \frac{1}{2} \left(1 - 3(t - t_0)^{-2n} \zeta_1 + 3\zeta_0\right). \label{eq:13}
\end{equation}
\begin{figure*}[htb]
\centerline{\includegraphics[scale=0.55]{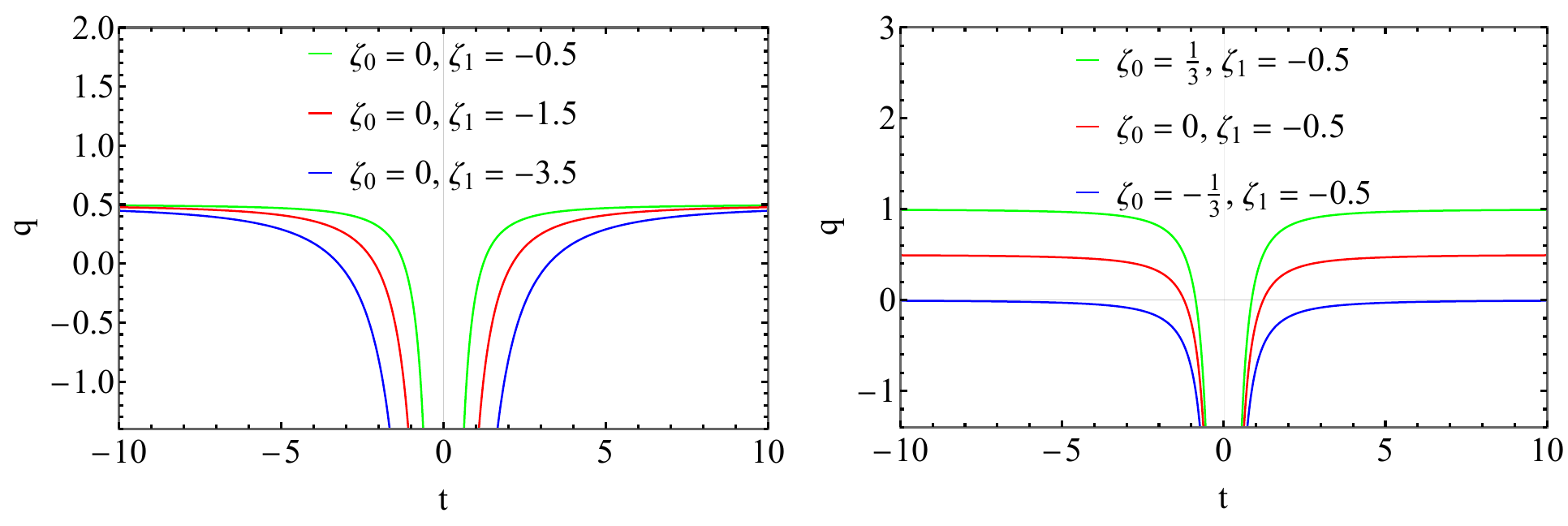}}
\caption{Variation of the deceleration parameter with cosmological time \( t \) for the Model. We have made the following assumptions for the plot: \( n = 1 \), \( t_0 = 0 \), and \( a_0 = 1 \). The unit of $t$ is in $Gyr$.}
\label{fig6}
\end{figure*}
\begin{table*}[htb]
\centering
\begin{tabular}{|c|c|c|}
\hline
\textbf{Cases} & \textbf{Conditions} & \textbf{Universe Scenario} \\
\hline
A    & \(H > 0,\ q > 0\)     & Expanding and Decelerating \\
B   & \(H > 0,\ q < 0\)     & Expanding and Accelerating \\
C  & \(H > 0,\ q = 0\)     & Expanding and Zero Deceleration / Constant Expansion \\
D  & \(H < 0,\ q > 0\)     & Contracting and Decelerating \\
E    & \(H < 0,\ q < 0\)     & Contracting and Accelerating \\
F  & \(H < 0,\ q = 0\)     & Contracting and Zero Deceleration / Constant Expansion \\
G  & \(H = 0,\ q = 0\)     & Static \\
\hline
\end{tabular}
\caption{Cosmological scenarios classification based on Hubble and deceleration parameter.}
\label{tab:1}
\end{table*}
\begin{table*}[htb]
\centering
\begin{tabular}{|c|c|c|}
\hline
\textbf{Case} & \textbf{Condition} & \textbf{Universe Scenario} \\
\hline
I & \(q < -1\) & Super Exponential Expansion \\
II & \(-1 \leq q < 0\) & Exponential Expansion (known as de-Sitter expansion) \\
III& \(q = 0\) & Expansion with Constant Rate \\
IV& \(-1 < q < 1\) & Accelerating Power Expansion \\
V& \(q > 0\) & Decelerating Expansion \\
\hline
\end{tabular}
\caption{Cosmological Scenarios for an expanding universe with different deceleration parameter values.}
\label{tab:2}
\end{table*}\\
We can see from Fig. \ref{fig6} that the deceleration parameter \( q \) described by Eq. (\ref{eq:13}) may reflect all kinds of conceivable universe situations in addition to the current expanding universe scenario. Furthermore, the deceleration parameter at the bouncing point \( t = 0 \) exhibits symmetrical behaviour, as seen in Fig. \ref{fig6}, and its fluctuation with the cosmic time is comparable to that found in other research  works \cite{ref82}. It should also  be noted that by only altering the model parameters \( \zeta_0 \) and \( \zeta_1 \) correspondingly, the deceleration parameter may be fitted with extreme precision to describe several potential universe situations. 

\subsection{Energy Conditions}\label{sub3}
In order to confirm the feasibility of the model, we will look at some energy conditions of well-known importance in this section. Current cosmology's cosmic acceleration may be predicted using some sets of energy requirements or conditions derived from the Friedmann equation. These energy requirements are essential to GR because they establish the premises for the existence of black holes and space-time singularity \cite{ref85}. Numerous authors from various backgrounds have worked on energy conditions. Subsequently well recognised energy conditions consist of the WEC (weak energy condition), NEC (null energy condition), DEC (dominant energy condition), and SEC (strong energy conditions), which are defined by \cite{ref86}:
\begin{itemize}
    \item Null Energy Condition (NEC) $\iff \rho + p \geq 0$
    \item Strong Energy Condition (SEC) $\iff \rho + 3p \geq 0$
    \item Dominant Energy Condition (DEC) $\iff \rho - p \geq 0$
    \item Weak Energy Condition (WEC) $\iff \rho \geq 0, \rho + p \geq 0$
\end{itemize}
For the model, we can define the pressure \( p \) as:
\begin{widetext}
\begin{equation}
    p = -\frac{4(1-2n)^2 (t - t_0)^{-2+2n} (\zeta_1 - (t - t_0)^{2n} \zeta_0)}{3\left(\zeta_1 - (t - t_0)^{2n} (1+\zeta_0) + 2n (t - t_0)^{2n} (1+\zeta_0)\right)^2}. \label{eq:15}
\end{equation}
\end{widetext}
\begin{figure*}[htb]
\centerline{\includegraphics[scale=0.75]{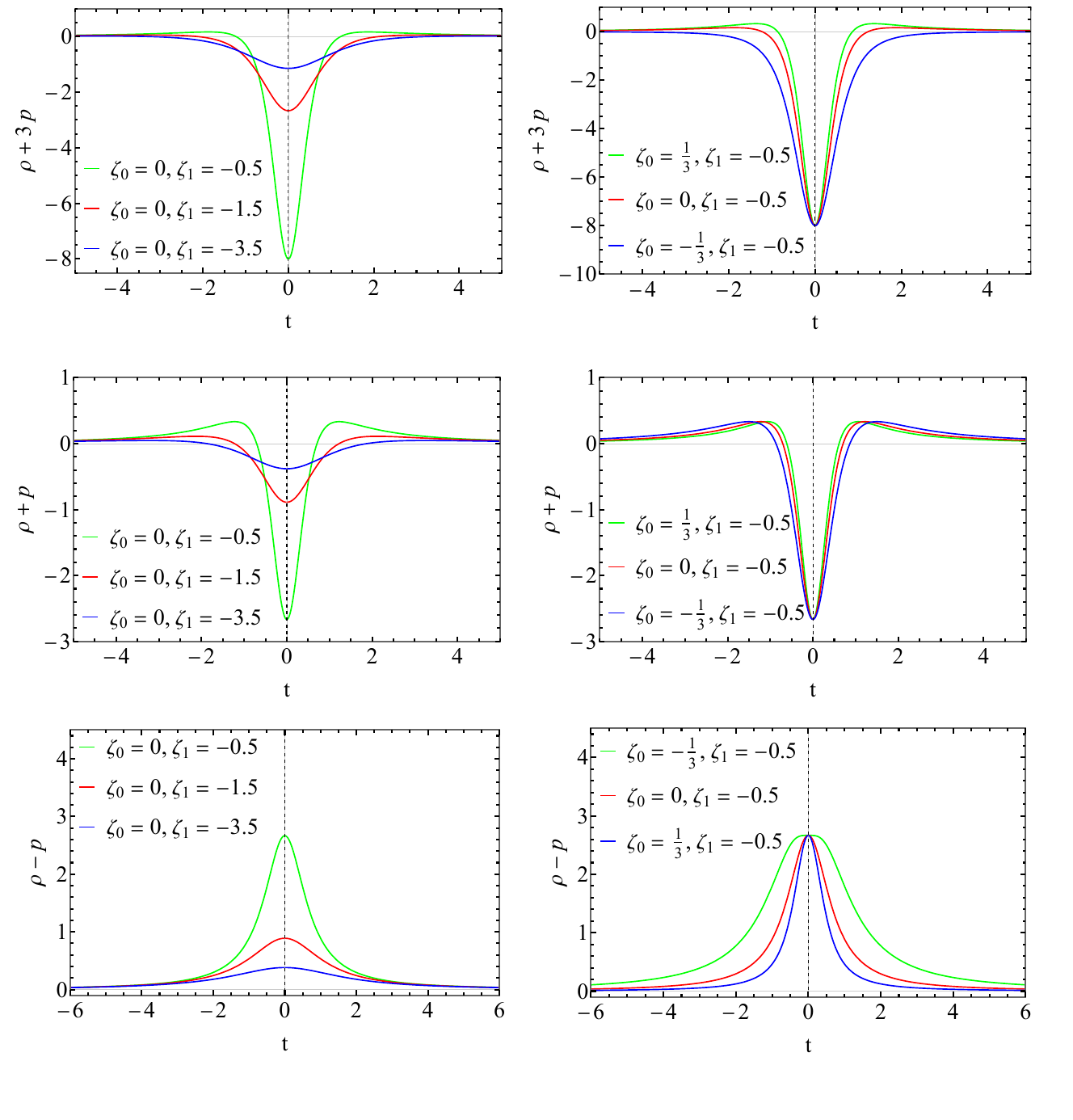}}
\caption{Evolution of the energy conditions with cosmological time. We have made the following assumptions for the plot: \( n = 1 \) and \( t_0 = 0 \). The unit of $t$ is in $Gyr$.}
\label{fig7}
\end{figure*}
Now using Eq. \ref{eq:11} and Eq. \ref{eq:15}, we evaluate $\rho + p$, $\rho + 3 p$ and $\rho - p$, as:
%\begin{widetext}
\begin{equation}
\rho + p = \frac{4 (1-2 n)^2 \left(t-t_0\right)^{2 n-2} \left(\zeta_1+\left(\zeta_0+1\right) \left(t-t_0\right)^{2 n}\right)}{3 \left(\zeta_1-\left(\zeta_0+1\right) (2 n-1) \left(t-t_0\right)^{2 n}\right)^2},
\end{equation}
\begin{equation}
\rho + 3p = \frac{4 (1-2 n)^2 \left(t-t_0\right){}^{2 n-2} \left(3 \zeta _1+\left(3 \zeta _0+1\right) \left(t-t_0\right){}^{2 n}\right)}{3 \left(\zeta _1-\left(\zeta _0+1\right) (2 n-1) \left(t-t_0\right){}^{2 n}\right)^2},
\end{equation}
\begin{equation}
\rho - p = -\frac{4 (1-2 n)^2 \left(\zeta _1+\left(\zeta _0-1\right) \left(t-t_0\right){}^{2 n}\right)}{3 \left(t-t_0\right){}^2 \left(-2 \left(\zeta _0+1\right) \zeta _1 (2 n-1)+\zeta _1^2 \left(t-t_0\right){}^{-2 n}+\left(\zeta _0+1\right){}^2 (1-2 n)^2 \left(t-t_0\right)^{2 n}\right)}.
\end{equation}
%\end{widetext}
As seen in Fig. \ref{fig7}, we provide a graphical illustration of them in relation to cosmic time. In technical terms, the violation of NEC represents the drop in energy density as the universe expands. The SEC violation is the cause of the universe's acceleration. Similarly, to realise a bounce, there must be a NEC violation. Furthermore, the SEC condition must be broken in order to depict a cosmic scenario where negative pressure prevails \cite{ref87}. The SEC have to be broken on a cosmic scale, with respect to the latest data on the expanding universe \cite{ref88,ref89,ref90}. The evolution of SEC demonstrates the rapidity at which the universe is expanding. The NEC and SEC are both violated at the bounce point, as can be seen from Fig. \ref{fig7}.  We also see a WEC violation, but there is no DEC violation. The bouncing scenario actually has to include a simple breach of the NEC as the universe expands, showing how the energy density is being depleted. The SEC violation is what is causing dark energy to exist, which meets the previously mentioned need for the model to match up with current measurements of the expanding universe. A non-singular bouncing universe is demonstrated by the fact that the NEC and SEC are clearly non-singular at the bouncing point, which resolves the early description's singularity issue.

\section{Stability Analysis}\label{sec4.1}
\begin{figure*}[htb]
\centerline{\includegraphics[scale=0.85]{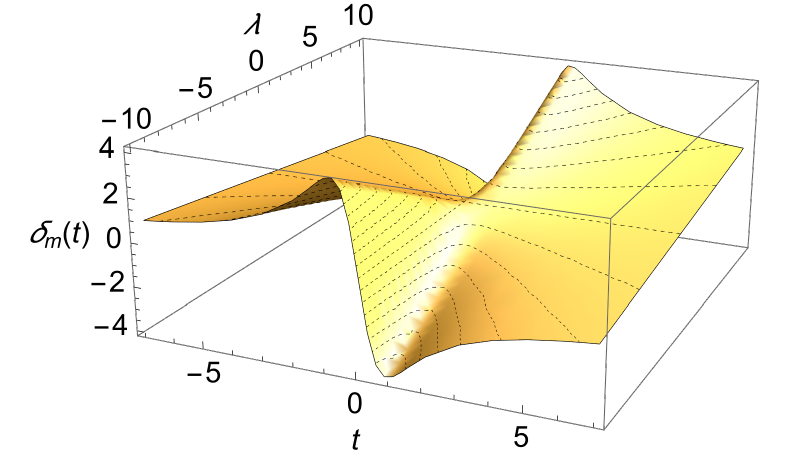}\hspace{0.3cm}\includegraphics[scale=0.85]{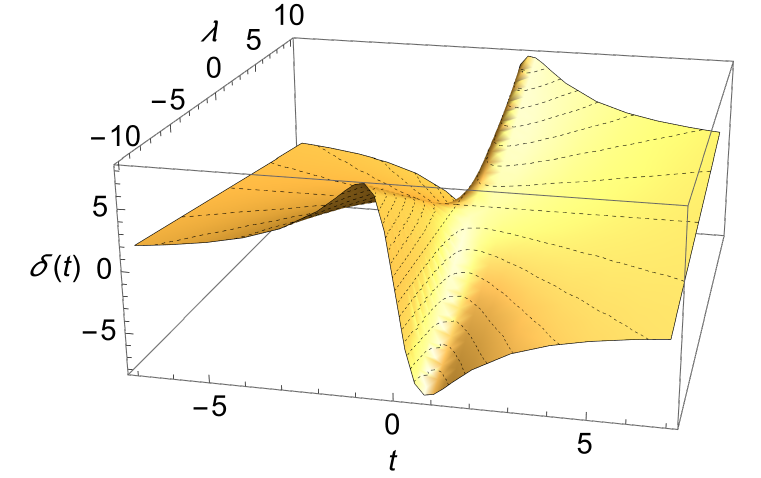}}
\caption{Variation of the perturbation terms $\delta_m(t)$ and $\delta(t)$ as a function of cosmic time t. We have made the following assumptions for the plot: $\zeta_0=0.5$, $\zeta_1=-1$, \( n = 1 \), and \( t_0 = 0 \).}
\label{figfan}
\end{figure*}

The stability of our cosmological model under homogeneous linear perturbations is discussed in this section. In particular, we define the first-order perturbation as follows \cite{ref90AN,ref90AN1} for both the energy density and the Hubble parameter:
\begin{equation}
H^{*}(t)=H(t)(1+\delta(t)),
\label{An1}
\end{equation} 
\begin{equation}
\rho^{*}(t)=\rho(t)(1+\delta_m(t)),
\label{An2}
\end{equation} 
where, $H^{*}(t)$ and $\rho^{*}(t)$ are the perturbed Hubble paramter and energy density, along with $\delta(t)$ and $\delta_m(t)$ as their corresponding perturbation terms. As we know, standard continuity equation in cosmology is defined as:
\begin{equation}
\dot{\rho}+3 H (\rho + p) = 0
\label{An3}
\end{equation}
Now, using Eqs. (\ref{An1}),(\ref{An2}), (\ref{eq:6}), and (\ref{eq:4}) in Eq. (\ref{An3}), we obtain:
\begin{equation}
\dot{\delta}_m(t)+3 H (1 + \zeta _0 + \zeta _1 \left(t-t_0\right){}^{-2 n}) \delta(t)= 0,
\label{An4}
\end{equation}
\begin{equation}
2\delta(t)=\delta_m(t).
\label{An5}
\end{equation}
Solving the above we evaluated $\delta(t)$ and $\delta_m(t)$ as:
\begin{equation}
\delta_m(t)=\frac{2 \lambda (t-t_0)}{\left(\left(\zeta _0+1\right) (2 n-1) \left(t-t_0\right){}^{2 n}-\zeta _1\right){}^{2 n-1}},
\label{An6}
\end{equation}
\begin{equation}
\delta(t)= \frac{\lambda (t-t_0)}{\left(\left(\zeta _0+1\right) (2 n-1) \left(t-t_0\right){}^{2 n}-\zeta _1\right){}^{2 n-1}},
\label{An7}
\end{equation}
where, $\lambda$ is some constant of integration. Fig. \ref{figfan} illustrates the evolution of the perturbation terms. We can see the evolution of both the perturbation terms are identical for the subsequent model parameters. At the early era both the perturbation terms rises before reaching some maximum value and then decreasing tending toward zero as the time evolves, similar to results found in other works \cite{ref90AN}. For a stable cosmic scenario, the perturbations should generally die out or at least stay limited to a finite as time increases. Both perturbation terms here approach zero after the bouncing point, indicating perturbations have stabilised and the universe has reverted to an isotropic and homogeneous condition. A desired characteristic for any viable cosmological model is stability under scalar perturbations, which is demonstrated by the stability analysis for the concerned model.
\section{Hubble Flow dynamics}\label{sec4}
\begin{figure*}[htb]
\centerline{
\includegraphics[width=.45\textwidth]{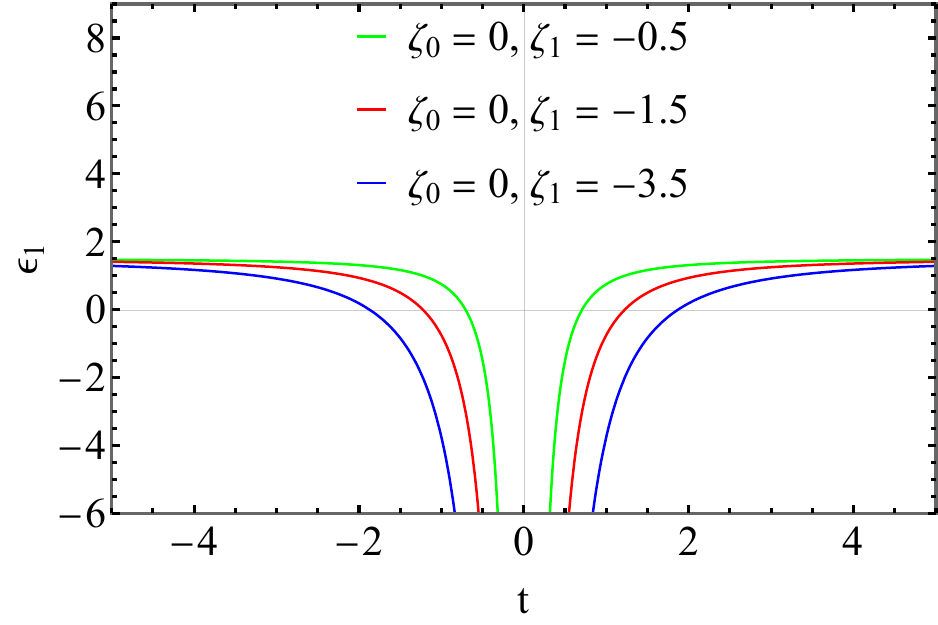}
\includegraphics[width=.45\textwidth]{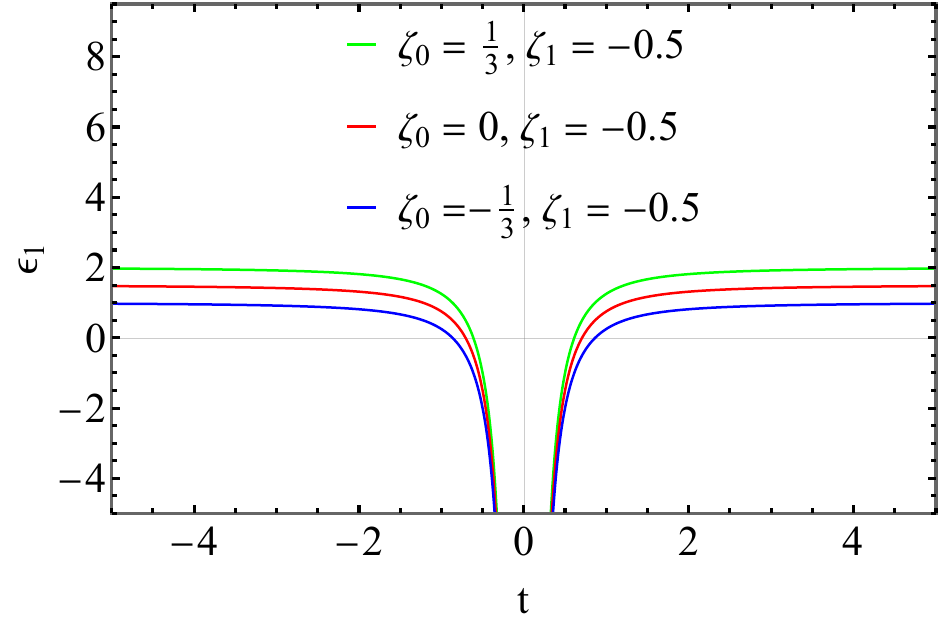}}
\centerline{
\includegraphics[width=.45\textwidth]{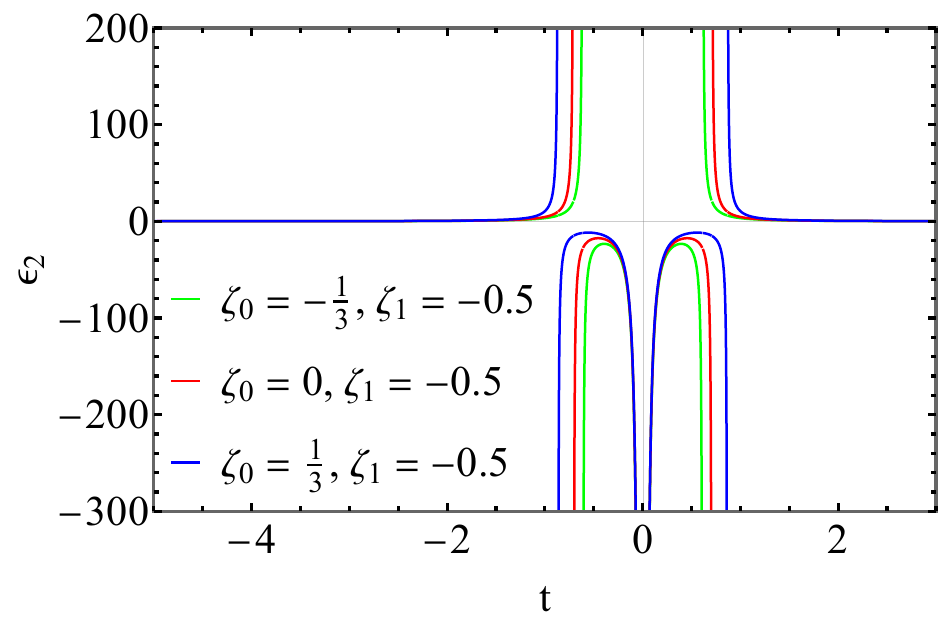}
\includegraphics[width=.45\textwidth]{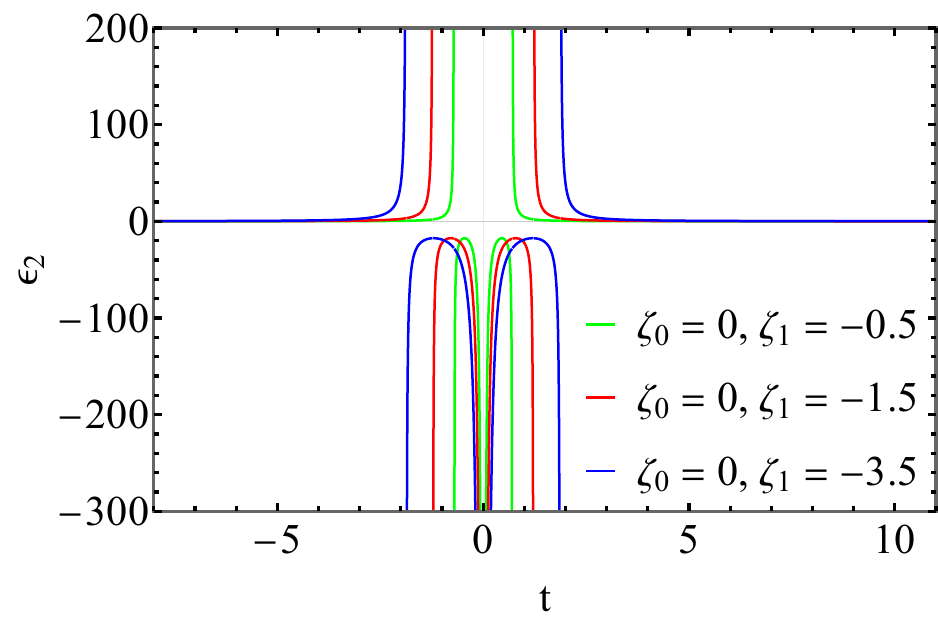}}
\centerline{
\includegraphics[width=.45\textwidth]{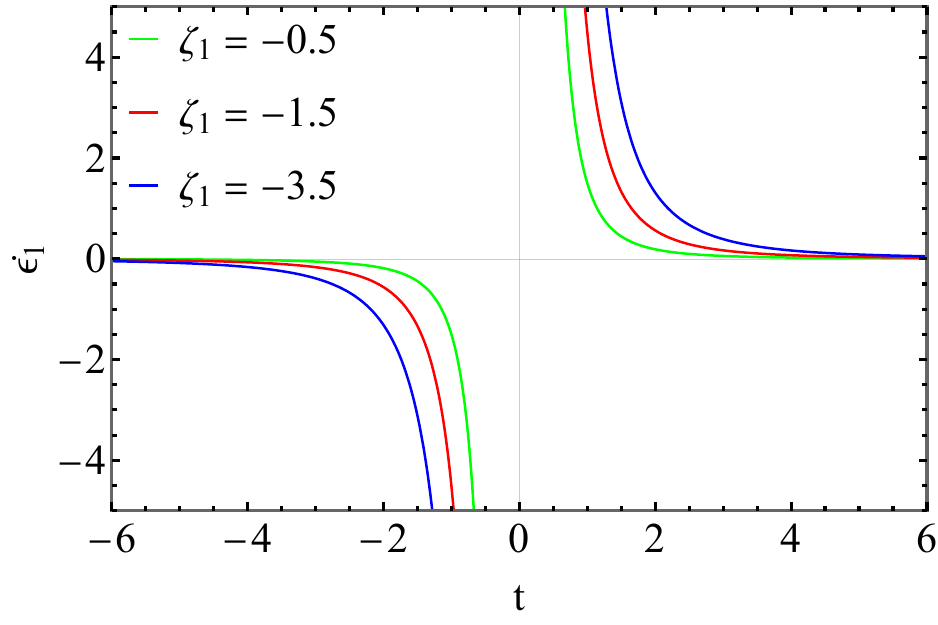}}
\caption{\label{fig8} Plot of Hubble flow parameter with respect to cosmological time $t$. We have made the following assumptions for the plot: $t_0=0$ and $n=1$. The unit of $t$ is in $Gyr$.}
\end{figure*}

As dictated by the inflation theory introduced in 1981, the early universe experienced a superluminal expansion as a result of its fast expansion following the Big Bang. In addition to solving the horizon problem, this idea is essential for comprehending the cosmology of the universe and early universe parameter explanation. Specifically, in this section we examine the model's ability to explain the fundamentals of the inflationary era or the transitions between the matter-bounce and inflation eras. We consider the following Hubble flow parameters \cite{ref91,ref92, ref93}, given by:
\begin{equation}
    \epsilon_{i+1} = \frac{d \ln{\epsilon_i}}{dN},
\end{equation}
where, \( N \) is the e-folding time scale. From it, we define the first two Hubble flow parameters \( \epsilon_1 \) and \( \epsilon_2 \) as:
\begin{equation}
    \epsilon_1 = 1 - \frac{\ddot{a}}{a H^2}, \quad \epsilon_2 = \frac{\dot{\epsilon_1}}{H \epsilon_1}.
\end{equation}
Since \( \ddot{a} > 0 \), \( \epsilon_1 \) must be smaller than 1 for inflation. It is evident that \( \epsilon_1 \) must assume the value of unity during a fading inflationary period. The expressions for \( \epsilon_1 \) and \( \epsilon_2 \) in terms of time evolution for our model are:
\begin{equation}
    \epsilon_1 = \frac{3}{2} \left( 1 + (t - t_0)^{-2n} \zeta_1 + \zeta_0 \right),
\label{17a}
\end{equation}
\begin{equation}
	\epsilon_2 = \frac{3 \zeta _1 n \left(t - t_0\right)^{-2 n}
	    \bigl(\zeta _0 - 2 (\zeta _0 + 1) n + \zeta _0 (t - t_0)^{-2 n} + 1\bigr)
	}{(2 n - 1)\,\bigl(\zeta _0 + \zeta _1 (t - t_0)^{-2 n} + 1\bigr)}
	\label{17b}
\end{equation}
In Fig. \ref{fig8}, we plot the Hubble flow parameters \( \epsilon_1 \) and \( \epsilon_2 \) with cosmological time using Eq. (\ref{17a}) and Eq. (\ref{17b}). Here, we observe a symmetric pattern of \( \epsilon_1 \) around the bouncing point or epoch. In the expanding phase, \( \epsilon_1 \) rises with growing cosmic time close to the bounce point or epoch, whereas in the contracting phase, \( \epsilon_1 \) falls with rising cosmic time. \( \epsilon_1 \ll 1 \), meeting the necessary requirement for the inflationary period, over the bounce epoch.  After that, subsequent values of the model parameters can also be seen to mediate the model to cross the boundary of \( \epsilon_1 = 1 \), indicating an exit from the inflationary era. Additionally, in the contracting phase near the bouncing epoch, we observe:
\begin{equation}
    \epsilon_1 < 0, \quad \epsilon_2 < 0, \quad \dot{\epsilon_1} < 0.
\end{equation}
Just after the bounce, in the expanding phase near the bouncing epoch, we have:
\begin{equation}
    \epsilon_1 < 0, \quad \epsilon_2 < 0, \quad \dot{\epsilon_1} > 0.
\end{equation}
These observations support the presence of inflation and demonstrate the capability of the model under our consideration to realize the inflationary era and exit from inflation. Similar analysis can also be found in Ref. \cite{ref94} and Ref.\cite{ref95} with identical results.
\section{Statefinder diagnostics}\label{sec5}
According to recent observational findings, hypothetical dark energy is causing the late-time universe to expand more quickly \cite{ref96}. Models are impacted by dark energy's effects on physical properties. For the analysis of universe expansion and dark energy properties in late-time universes, geometrical parameters derived from the space-time metric are more relevant and invariant to uncertainty. A simple diagnostic pair used to study dark energy behaviour without relying on particular models is the statefinder diagnostic pair \((r, s)\) along with the \((r, q)\) pair \cite{ref97}. The scale factor can be used directly to build the statefinder parameters \( r \) and \( s \), defined by:
\begin{equation}
    (r, s) = \left(\frac{\dddot{a}}{a H^3}, \frac{r-1}{3(q - 0.5)}\right),
\end{equation}
where, \( \dddot{a} \) denotes the third-order time derivative of the scale factor \( a \), with \( q \) as the deceleration parameter, provided \( q \neq \frac{1}{2} \). 
\begin{table*}[htb]
    \centering
    \begin{tabular}{|c|c|}
        \hline
        \textbf{Model} & \(\mathbf{(r, s)}\) \\
        \hline
        \(\Lambda\)CDM & $(1, 0)$ \\
        SCDM & $(1, 1)$ \\
        Quintessence & $(<1, >0)$ \\
        Chaplygin Gas & $(>1, <0)$ \\
        \hline
    \end{tabular}
    \caption{Statefinder diagnostic values for different models.}
\end{table*}
For our model, we have the values of \( r \), $q$ and \( s \) as:
\begin{widetext}
\begin{equation}
    r = \frac{1}{2} \left(9 \zeta _0^2+9 \zeta _0+\frac{9 \zeta _1^2 (n-1) \left(t-t_0\right){}^{-4 n}}{2 n-1}+9 \zeta _1 \left(\zeta _0 (n+2)+n+1\right) \left(t-t_0\right){}^{-2 n}+2\right),
\end{equation}

\begin{equation}
    q = \frac{1}{2} \left(3 \zeta _0+3 \zeta _1 \left(t-t_0\right){}^{-2 n}+1\right),
\end{equation}

\begin{equation}
    s = \frac{3 \left(\zeta _0^2+\zeta _0+\frac{\zeta _1^2 (n-1) \left(t-t_0\right){}^{-4 n}}{2 n-1}+\zeta _1 \left(\zeta _0 (n+2)+n+1\right) \left(t-t_0\right){}^{-2 n}\right)}{2 \left(1.5 \zeta _0+1.5 \zeta _1 \left(t-t_0\right){}^{-2 n}\right)}.
\end{equation}
\end{widetext}
\begin{figure*}[htb]
\centerline{\includegraphics[scale=0.5]{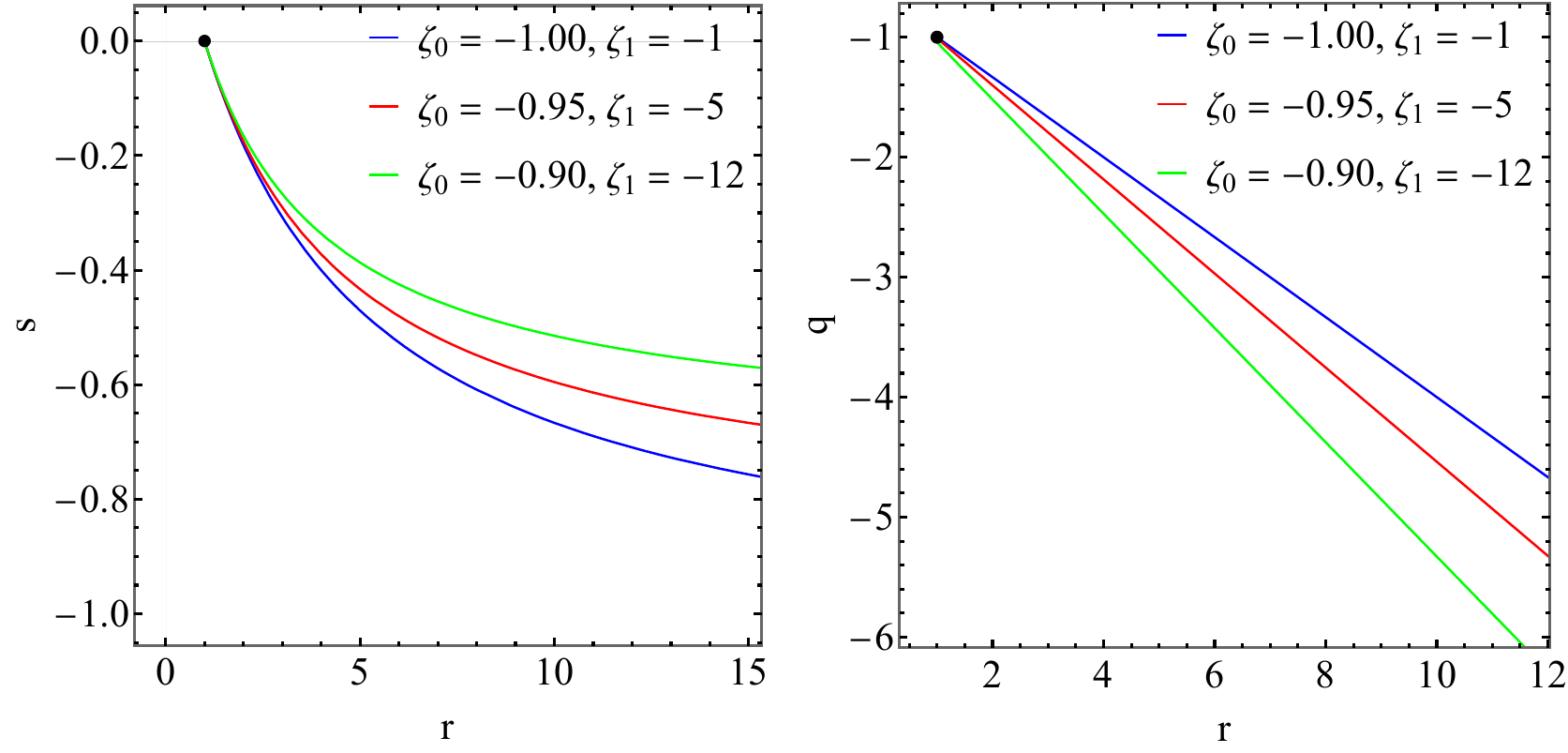}}
\caption{Plot for the statefinder diagnostics in the $(r,s)$ and $(r,q)$ plane for the model. For the plot we assumed $a_0=1$, $t_0=0$ and $n=1$. Here, the black dot in the plane of  $(r,s)$ and $(r,q)$ denote the \(\Lambda\)CDM point where $(r,s)=(1,0)$ and $(r,q)=(1,-1)$.}
\label{fig9}
\end{figure*}
As seen in Fig. \ref{fig9}, we construct a parametric plot between \( r \) and \( s \), as well as \( r \) and \( q \), using the equations for \( r \), \( s \), and \( q \) as derived. The model's temporal development in the \( r - s \) and \( r - q \) phase planes is represented by the arrow's direction. As a result, we can see that the universe in our model shows transition from an initial state to the usual \(\Lambda\)CDM phase. For various parameter values, the dark energy behaviour of our considered model is Chaplygin Gas-type. In every instance, nevertheless, the model eventually transitions into the typical \(\Lambda\)CDM phase.

\section{Comparison with Observational Data}\label{sec6}
The ability of a cosmological model to describe the current universe using certain observational datasets is a key indicator of its its viability.  We examine the Hubble parameter, distance modulus, and Hubble distance evolution with redshift in this section in an attempt to determine the extent to which our model can account for observations related to the late-time universe. We utilize statistical parameters such $R^2$, $\chi^2_{\min}$, AIC (Akaike Information Criterion), and BIC (Bayesian Information Criterion) to compare the model with $\Lambda$CDM for various datasets (refer to Ref. \cite{refS66}, Ref. \cite{refS68}, and Ref. \cite{refS64} for review). The model that best fits the data statistically is thought to be the one with the lowest $\chi^2_{\min}$, AIC, and BIC values.  Likewise, the model with the $R^2$ value closest to 1 is the one that the data favors the most. The only two of these statistical criteria that do not take into account the number of free parameters in the model under study are $R^2$ and $\chi^2_{\min}$. Both AIC and BIC are model selection criteria that limit models with higher complexity or free parameters, although they work in different ways. AIC promotes the models that balance goodness of fit and model complexity, but BIC is perceived as being stricter and favoring simpler models, especially when working with larger datasets.  Because of its capability to penalise complexity, which increases with sample size, BIC is less likely than AIC to select the model's over parametrization.  The AIC or BIC difference between the models is calculated (i.e., $\Delta X = \Delta \mathrm{AIC}$ or $\Delta X = \Delta \mathrm{BIC}$) in order to compare them. The evidence is \textit{weak} and it is impossible to judge whether model is superior if $0 \leq \Delta X \leq 2$ or $-2 \leq \Delta X < 0$.  Evidence is \textit{positive} in support of the model with the lower value if $2 < \Delta X \leq 6$ or $-6 \leq \Delta X < -2$.  Evidence is considered to be \textit{strong} if $6 < \Delta X \leq 10$ or $-10 \leq \Delta X < -6$.

\subsection{Hubble Parameter Vs Redshift}
\begin{figure*}[htb]
\centerline{\includegraphics[scale=0.55]{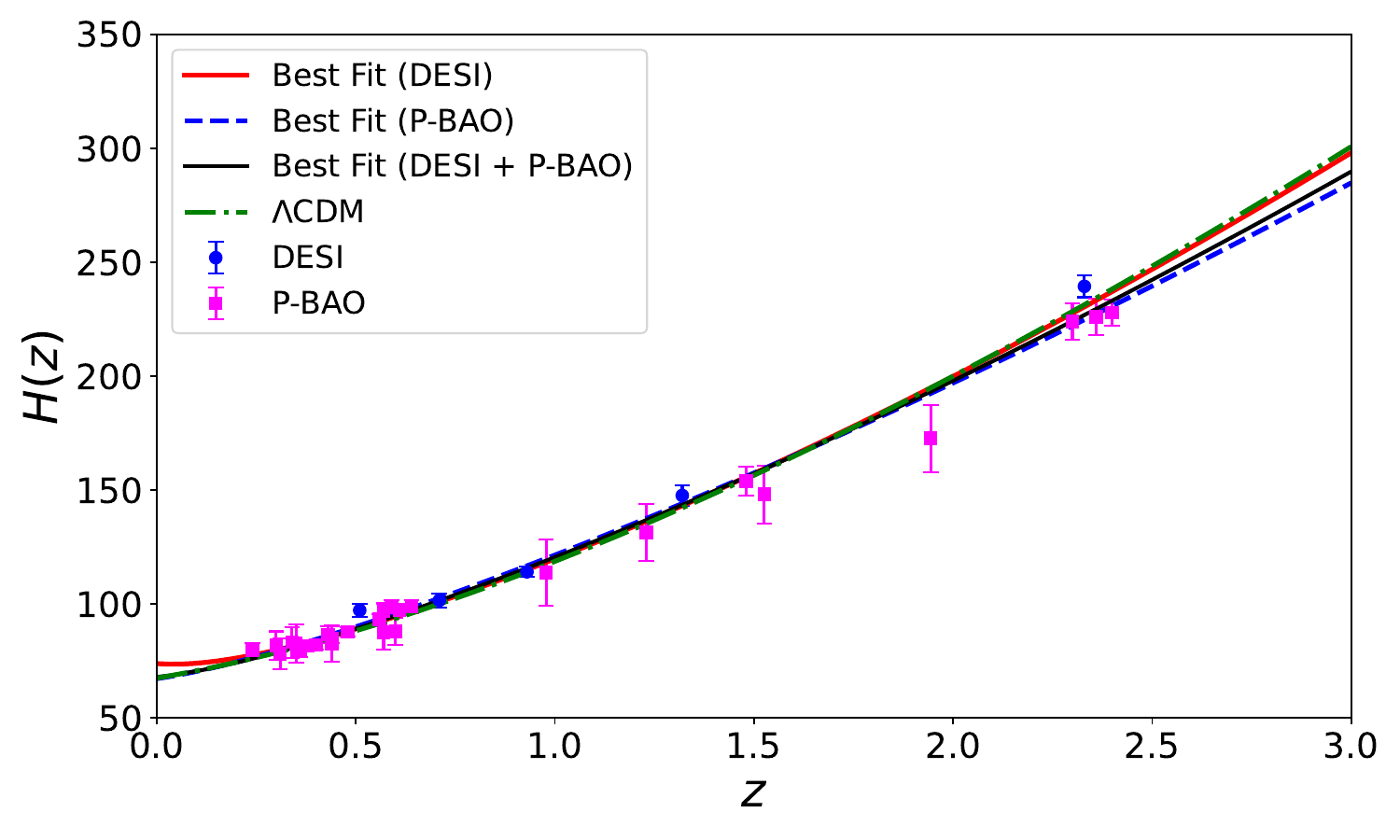}}
\caption{Plot of $H(z)$ vs $z$ for the best fit of the model parameters against the observational data. Here, a comparison is made with the $\Lambda$CDM model.}
\label{fig10}
\end{figure*}

\begin{figure*}[htb]
\centerline{\includegraphics[scale=0.6]{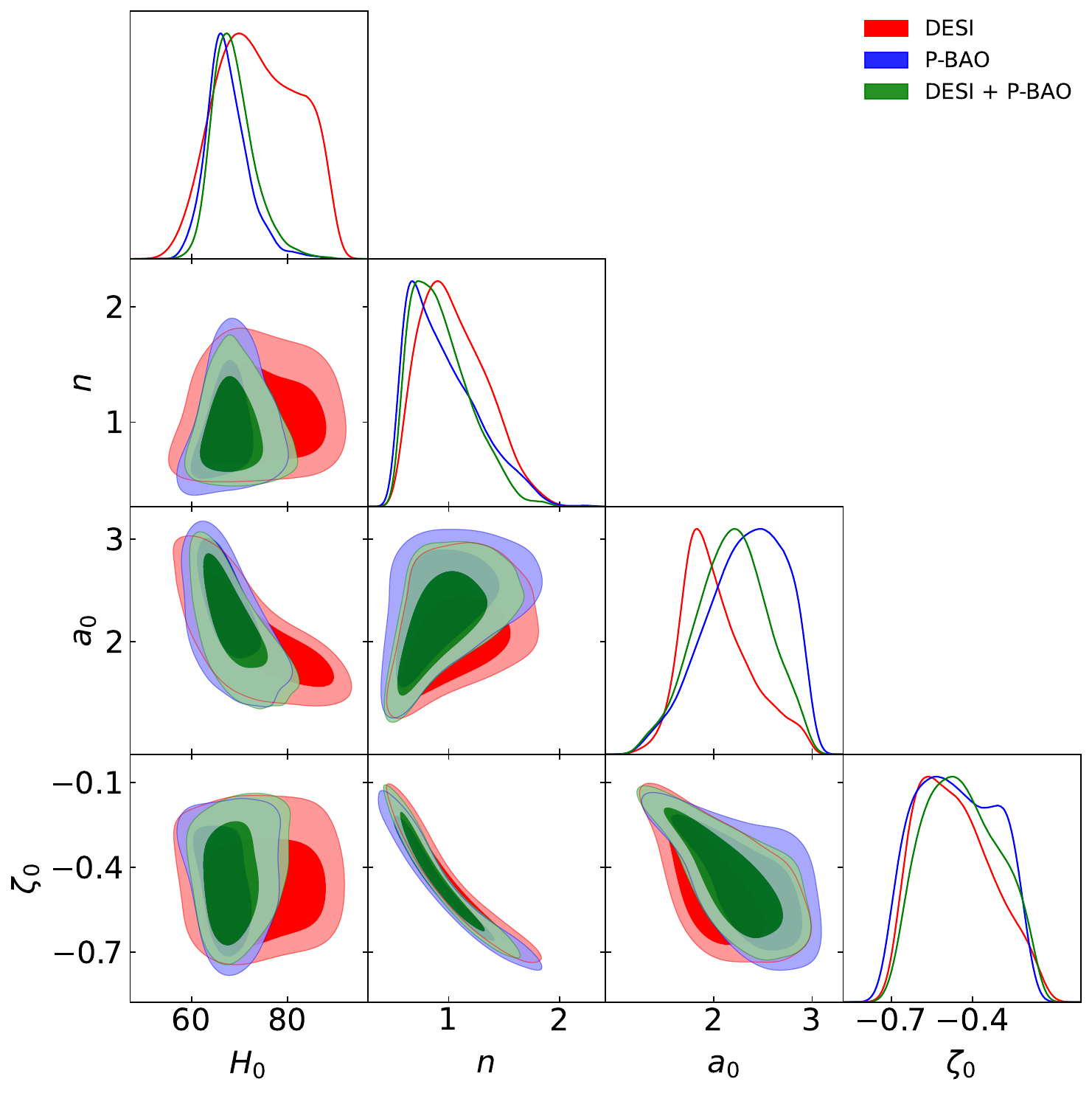}}
\caption{2-d contour sub-plot for the parameters $H_0$, $a_0$, $n$ and $\zeta_0$ with 1-$\sigma$ and 2-$\sigma$ errors (showing the 68\% and 95\% c.l.) for $H(z)$ vs $z$. }
\label{fig10_2}
\end{figure*}

The current expansion rate of the the universe is determined using the observational data like the Hubble data (OHD). The primary purpose of these data is to calculate the Hubble parameter $H(z)$ as a function of Redshift $z$, independent of the cosmological model.  As shown in Table \ref{table:4} for Hubble parameter $H(z)$, we consider two datasets from the Dark Energy Spectroscopic Instrument collaboration (DESI) and previous Baryonic Acoustic Oscillations (BAO), or what we refer to in the research as P-BAO from observations such as SDSS and WiggleZ. Now to derive the Hubble parameter in terms of redshift, we express Eq. (\ref{eq:7}) in terms of the scale factor $a$ by substituting the time derivative $\frac{d}{dt}$ to $H\frac{d}{d ln a}$, and integrating it we obtain:
\begin{equation}
H=\left(\frac{a}{C_3}\right)^{\alpha (t-t_0)^{-2n}- \beta},
\label{28}
\end{equation}
where, $C_3$ is some constant of integration. Now using Eq. (\ref{eq:10}) we can obtain:
\begin{equation}
(t-t_0)^{2n}=\frac{\left(\frac{a (\alpha  n)^{\frac{1}{2 \beta  n}}}{\text{a0}}\right)^{2 \beta  n}-\alpha  n}{\beta  n (2 n-1)}
\label{29}
\end{equation}
\begin{table}[h!]
\centering
\begin{tabular}{|ccc|c||ccc|c|}
\hline
\multicolumn{4}{|c||}{\textbf{DESI}} & \multicolumn{4}{c|}{\textbf{P-BAO}} \\
\hline
$z$ & $H(z)$ & $\sigma_H$ & Ref & $z$ & $H(z)$ & $\sigma_H$ & Ref \\
\hline
0.51 & 97.21 & 2.83 & \cite{ref98} & 0.24 & 79.69 & 2.99 & \cite{d113} \\
0.71 & 101.57 & 3.04 & \cite{ref98} & 0.30 & 81.70 & 6.22 & \cite{d114} \\
0.93 & 114.07 & 2.24 & \cite{ref98} & 0.31 & 78.17 & 6.74 & \cite{d115} \\
1.32 & 147.58 & 4.49 & \cite{ref98} & 0.34 & 83.17 & 6.74 & \cite{d113} \\
2.33 & 239.38 & 4.80 & \cite{ref98} & 0.35 & 82.70 & 8.40 & \cite{d116} \\
 &  &  &  & 0.36 & 79.93 & 3.39 & \cite{d115} \\
 &  &  &  & 0.38 & 81.50 & 1.90 & \cite{d5} \\
 &  &  &  & 0.40 & 82.04 & 2.03 & \cite{d115} \\
 &  &  &  & 0.43 & 86.45 & 3.68 & \cite{d113} \\
 &  &  &  & 0.44 & 82.60 & 7.80 & \cite{d74} \\
 &  &  &  & 0.44 & 84.81 & 1.83 & \cite{d115} \\
 &  &  &  & 0.48 & 87.79 & 2.03 & \cite{d115} \\
 &  &  &  & 0.56 & 93.33 & 2.32 & \cite{d115} \\
 &  &  &  & 0.57 & 87.60 & 7.80 & \cite{d10} \\
 &  &  &  & 0.57 & 96.80 & 3.40 & \cite{d117} \\
 &  &  &  & 0.59 & 98.48 & 3.19 & \cite{d115} \\
 &  &  &  & 0.60 & 87.90 & 6.10 & \cite{d74} \\
 &  &  &  & 0.61 & 97.30 & 2.10 & \cite{d5} \\
 &  &  &  & 0.64 & 98.82 & 2.99 & \cite{d115} \\
 &  &  &  & 0.978 & 113.72 & 14.63 & \cite{d118} \\
 &  &  &  & 1.23 & 131.44 & 12.42 & \cite{d118} \\
 &  &  &  & 1.48 & 153.81 & 6.39 & \cite{d79} \\
 &  &  &  & 1.526 & 148.11 & 12.71 & \cite{d118} \\
 &  &  &  & 1.944 & 172.63 & 14.79 & \cite{d118} \\
 &  &  &  & 2.30 & 224.00 & 8.00 & \cite{d119} \\
 &  &  &  & 2.36 & 226.00 & 8.00 & \cite{d120} \\
 &  &  &  & 2.40 & 227.80 & 5.61 & \cite{d121} \\
\hline
\end{tabular}
\caption{Observed Hubble parameter $H(z)$ (in units of $km s^{-1} Mpc^{-1}$) and their uncertainties at redshift $z$ form the DESI and P-BAO datasets.}
\label{table:4}
\end{table}
On using Eq. (\ref{29}) in Eq. (\ref{28}), along with the relation $a=\frac{1}{1+z}$ where $a$ is the scale factor and $z$ is the redshift, we get:
\begin{equation}
H(z)=\left(\frac{1}{C_3 (z+1)}\right)^{\beta  \left(-\frac{\alpha  n (2 n-1)}{\alpha  n-\left(\frac{{\left( \alpha  n\right)}^{\frac{1}{2 \beta  n}}}{a_0 (z+1)}\right)^{2 \beta  n}}-1\right)}.
\label{30}
\end{equation}
Setting $H(z)=H_0$ at $z=0$ (Hubble constant at present epoch) and using Eq. (\ref{30}) along with aforementioned definition \(\alpha = -\frac{3}{2} \zeta_1\) and \(\beta = \frac{3}{2} (1 + \zeta_0)\) we obtain the final expression for Hubble parameter in terms of redshift:
\begin{widetext}
\begin{equation}
H(z)=\left( H_0^{\frac{2 \left(1-a_0^{3 \left(\zeta _0+1\right) n}\right)}{3 \left(\zeta _0+1\right) \left(2 n a_0^{3 \left(\zeta _0+1\right) n}-1\right)}}(1+z)\right){}^{\frac{3 \left(\zeta _0+1\right) \left(2 n \left(a_0 (z+1)\right){}^{3 \left(\zeta _0+1\right) n}-1\right)}{2 \left((\text{a0} (z+1))^{3 \left(\zeta _0+1\right) n}-1\right)}}.
\label{31}
\end{equation}
\end{widetext}
Now, using the datasets given in Table \ref{table:4} we perform statistical analysis using Markov Chain Monte Carlo (MCMC) method for our model, to find the best fit parameter values. To find the mean values of the parameters $H_0$, $a_0$, $n$ and $\zeta_0$ we have used the chi-squared function defined as:
\begin{equation}
\chi^2_H(H_0,n,a_0,\zeta_0) = \sum_{i=1}^{N} \frac{\left[ H_{th}(z_i, H_0,n,a_0,\zeta_0) - H_{obs}(z_i) \right]^2}{\sigma_H(z_i)^2},
\label{32}
\end{equation}
where, $\sigma_H(z_i)$ denotes the standard error in the observed value of $H(z_i)$ at redshift $z_i$, $H_{th}$ represents the theoretical value of the Hubble parameter, and $H_{obs}$ represents the observed value. The MCMC is initially conducted using the DESI dataset alone, followed by the previous BAO dataset, and finally, a combination of both datasets. Furthermore, we contrast the outcomes of the model with the $\Lambda$CDM model, as seen in Fig. \ref{fig10}. Table \ref{table:5} provides the $R^2$, $\chi^2_{\min}$, AIC and BIC values, which we compute to compare the goodness of fit of the model for each case with $\Lambda$CDM. Here, Table \ref{table:6} displays the optimal fitting parameters found for each scenario. Fig. \ref{fig10_2} shows the 2-d contour sub-plot for the parameters $H_0$, $a_0$, $n$ and $\zeta_0$ with 1-$\sigma$ and 2-$\sigma$ errors (showing the 68\% and 95\% c.l.) for $H(z)$ vs $z$. 

\begin{table}[htb]
\centering
\begin{tabular}{|l|c|c|c|}
\hline
\textbf{Statistical Parameters} & \textbf{DESI} & \textbf{P-BAO} & \textbf{DESI + P-BAO} \\
\hline
\multicolumn{4}{|c|}{\textbf{Model}} \\
\hline
$R^2$ Score        & 0.9896 & 0.9883 & 0.9858 \\
$\chi^2_{\min}$    & 8.65  & 12.50  &  21.15  \\
AIC                & 16.65  & 20.50  & 29.15  \\
BIC                & 15.07  & 25.68  & 35.01  \\
\hline
\multicolumn{4}{|c|}{\textbf{$\Lambda$CDM}} \\
\hline
$R^2$ Score        & 0.9878 & 0.9830 & 0.9846 \\
$\chi^2_{\min}$    & 13.58  & 21.03  & 34.60  \\
AIC                & 19.58  & 27.03  & 40.60  \\
BIC                & 18.41  & 30.91  & 45.00  \\
\hline
\multicolumn{4}{|c|}{\textbf{$\Delta$AIC and $\Delta$BIC}} \\
\hline
$\Delta$AIC        & 2.93   & 6.53   & 11.45   \\
$\Delta$BIC        & 3.34   & 5.23   & 9.99  \\
\hline
\end{tabular}
\caption{Statistical comparison between the model and $\Lambda$CDM for different datasets of $H(z)$ vs $z$. $\Delta$AIC and $\Delta$BIC represent the difference ($\Lambda$CDM $-$ Model) for AIC and BIC.}
\label{table:5}
\end{table}

\begin{table*}[ht]
\centering
\begin{tabular}{|l|c|c|c|c|}
\hline
\textbf{Data Set} & \boldmath$H_0$ & \boldmath$n$ & \boldmath$a_0$ & \boldmath$\zeta_0$ \\
\hline
\textbf{DESI} & $73.9780^{+10.1873}_{-8.5336}$  & $1.0085^{+0.3461}_{-0.2699}$ & $1.9485^{+0.4346}_{-0.2505}$ & $-0.4771^{+0.1682}_{-0.1300}$ \\
\textbf{P-BAO} & $66.8925^{+4.8217}_{-3.4453}$   & $0.8651^{+0.3765}_{-0.2470}$ & $2.3183^{+0.4117}_{-0.4339}$ & $-0.4504^{+0.1709}_{-0.1540}$ \\
\textbf{DESI + P-BAO}   & $68.2019^{+4.6316}_{-3.3796}$   & $0.9201^{+0.3821}_{-0.2544}$ & $2.2135^{+0.3809}_{-0.4058}$ & $-0.4631^{+0.1780}_{-0.1501}$ \\
\hline
\end{tabular}
\caption{Best-fit parameters with $1\sigma$ uncertainties for each dataset of $H(z)$.}
\label{table:6}
\end{table*}
The model's $\chi^2_{\min}$, AIC, and BIC values are lower than those of $\Lambda$CDM for each dataset of $H(z)$ vs $z$, according to the results. Additionally, in every instance, both $\Delta$AIC and $\Delta$BIC are $>2$, suggesting that the model is supported by positive evidence in comparison to $\Lambda$CDM. The success of the model as a whole is further demonstrated by the fact that the $R^2$ values in each case indicate a somewhat greater goodness of fit with observational data than the $\Lambda$CDM model. Hence, findings show that the model can adequately describe the DESI and P-BAO datasets as well as their combined versions for the evolution of the Hubble parameter $H(z)$ with redshift $z$. 
\subsection{Distance Modulus Vs Redshift}
\begin{figure*}[htb]
\centerline{\includegraphics[scale=0.5]{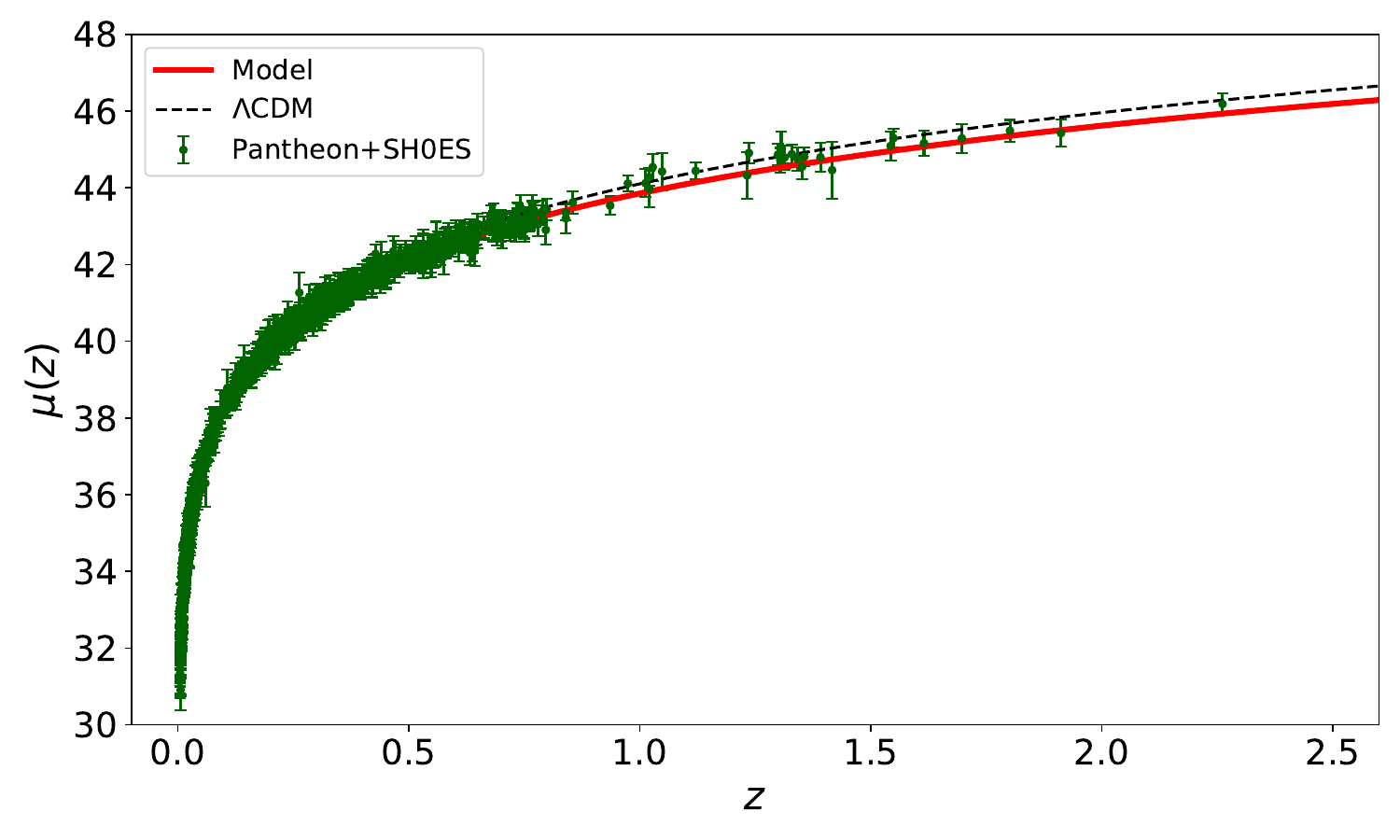}}
\caption{Plot of $\mu(z)$ vs $z$ for the best fit of the model parameters against the observational data. Here, a comparison is made with the $\Lambda$CDM model.}
\label{fig11}
\end{figure*}

\begin{figure*}[htb]
\centerline{\includegraphics[scale=0.5]{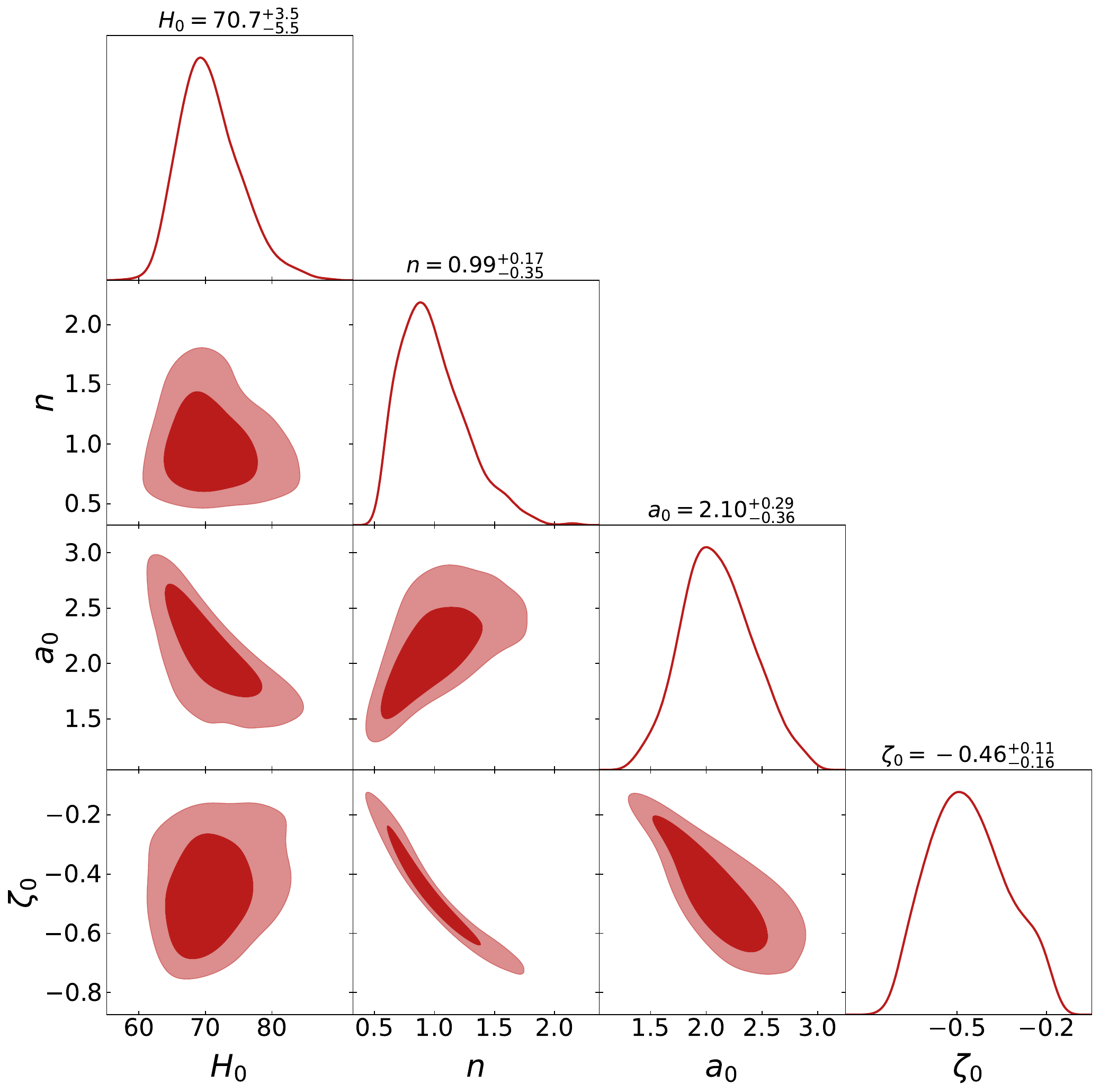}}
\caption{2-d contour sub-plot for the parameters $H_0$, $a_0$, $n$ and $\zeta_0$ with 1-$\sigma$ and 2-$\sigma$ errors (showing the 68\% and 95\% c.l.), with best fitting parameter values for $\mu(z)$ vs $z$.}
\label{fig11_2}
\end{figure*}

Here we shall investigate the  redshift-luminosity distance modulus relationship for our model. An efficient observational technique for examining the late-time development of the universe is the redshift-luminosity distance connection. In particular, we consider 1590 data from the Pantheon+ compilation \cite{refd9044,refd9045} with $z > 0.01$, supplemented with 42 SNeIa calibrated by SH0ES Cepheids of the host galaxies  \cite{refd9046}. We express the luminosity distance $d_L$ in terms of the redshift to calculate the distance modulus $\mu$ by employing the relation \cite{ref2}:
\begin{equation}
\mu = 5 \log_{10} d_L(z) + \mu_0,
\end{equation}
where, $\mu_0 = 5 \log_{10} \left( H_0^{-1}/ Mpc \right) + 25$ with $H_0$ being the dimensionless Hubble parameter. The luminosity distance in terms of redshift $z$ is given by:
\begin{equation}
d_L = c (1 + z) \int_0^z \frac{dz'}{H(z')}.
\label{hoi}
\end{equation}
To find the mean values of the parameters $H_0$, $a_0$, $n$ and $\zeta_0$ we have used the chi-squared function defined as:
\begin{widetext}
\begin{equation}
\chi^2_{SN}(H_0,n,a_0,\zeta_0) = \sum_{i=1}^{N} \frac{\left[ \mu_{th}(z_i, H_0,n,a_0,\zeta_0) - \mu_{obs}(z_i) \right]^2}{\sigma_\mu(z_i)^2},
\label{32a}
\end{equation}
\end{widetext}
where, $\sigma_\mu(z_i)$ denotes the standard error in the observed value of $\mu(z_i)$ at redshift $z_i$, $\mu_{th}$ represents the theoretical value of the distance modulus, and $H_{obs}$ represents the observed value. By performing MCMC the best fitting model parameters are obtained and results are compared with $\Lambda$CDM model. The Fig. \ref{fig11} shows effective fit of our model to the observational data. The $R^2$ value in this case also is greater then the $\Lambda$CDM model, indicating that our model fits the data better. Table \ref{table:9a} shows the values of $\chi^2_{\min}$, AIC, and BIC values are lower than those of $\Lambda$CDM for the dataset of $\mu(z)$ vs $z$. Furthermore, both $\Delta$AIC and $\Delta$BIC are sufficiently high, suggesting that the model is supported by very strong evidence in comparison to $\Lambda$CDM. Fig. \ref{fig11_2} shows the 2-d contour sub-plot for the parameters $H_0$, $a_0$, $n$ and $\zeta_0$ with 1-$\sigma$ and 2-$\sigma$ errors (showing the 68\% and 95\% c.l.), with best fitting parameter values for $\mu(z)$ vs $z$.
\begin{table*}[htb]
\centering
\begin{tabular}{|l|c|c|c|c|c|c|}
\hline
\textbf{Description} & $R^2$ Score & $\chi^2_{\min}$ & AIC & BIC & $\Delta$AIC & $\Delta$BIC \\
\hline
Model & 0.99708 & 887.411 & 895.411 & 917.117 & 306.544 & 295.691  \\
$\Lambda$CDM & 0.99651 & 1197.955 & 1201.955 & 1212.808 & - & - \\
\hline
\end{tabular}
\caption{Statistical comparison between the model and $\Lambda$CDM for the dataset of $\mu(z)$ vs $z$. $\Delta$AIC and $\Delta$BIC represent the difference ($\Lambda$CDM $-$ Model) for AIC and BIC.}
\label{table:9a}
\end{table*}

\subsection{Hubble distance Vs Redshift}
\begin{figure*}[htb]
\centerline{\includegraphics[scale=0.47]{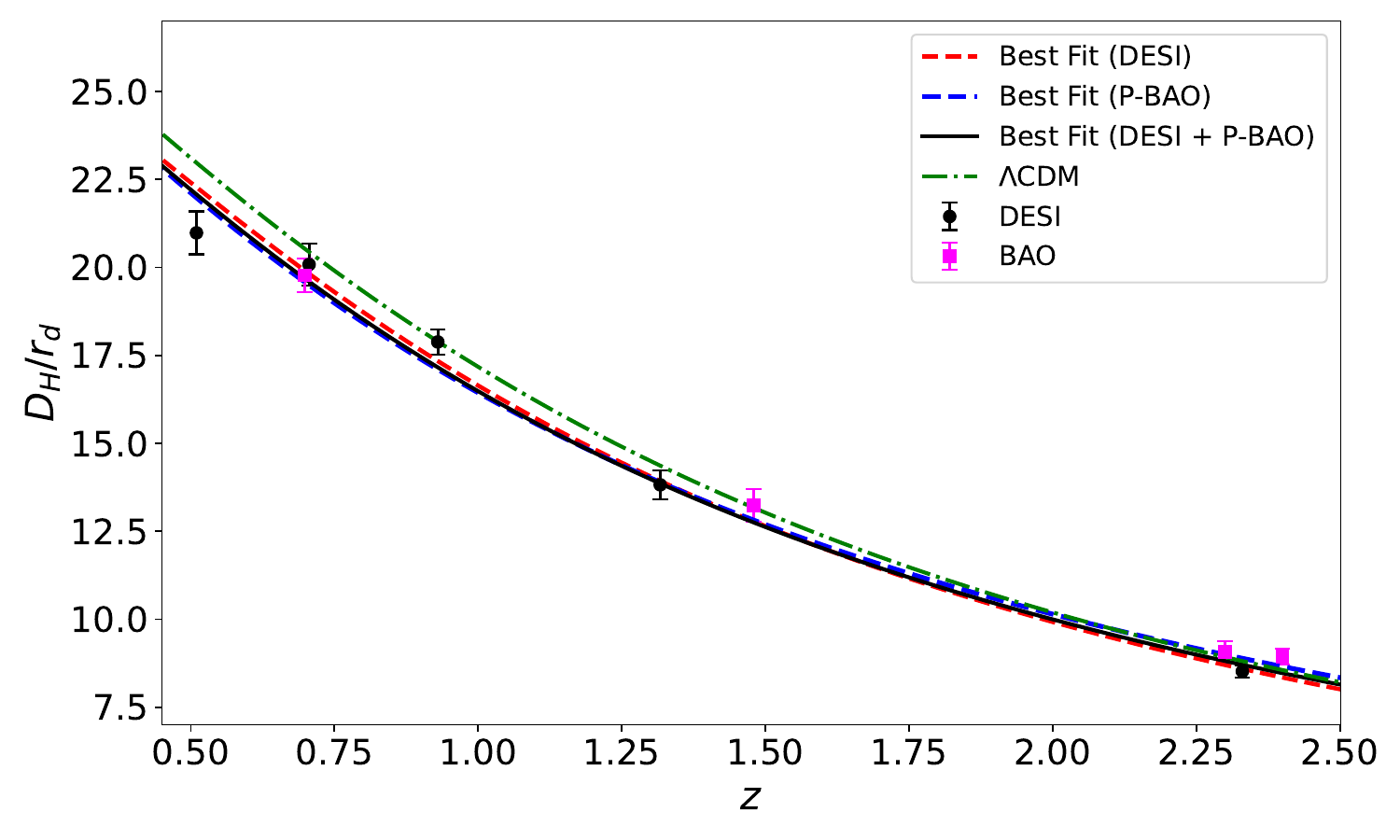}}
\caption{Plot of $D_H(z)$ vs $z$ for the best fit of the model parameters against the observational data. Here, a comparison is made with the $\Lambda$CDM model.}
\label{fig12}
\end{figure*}

\begin{figure*}[htb]
\centerline{\includegraphics[scale=0.6]{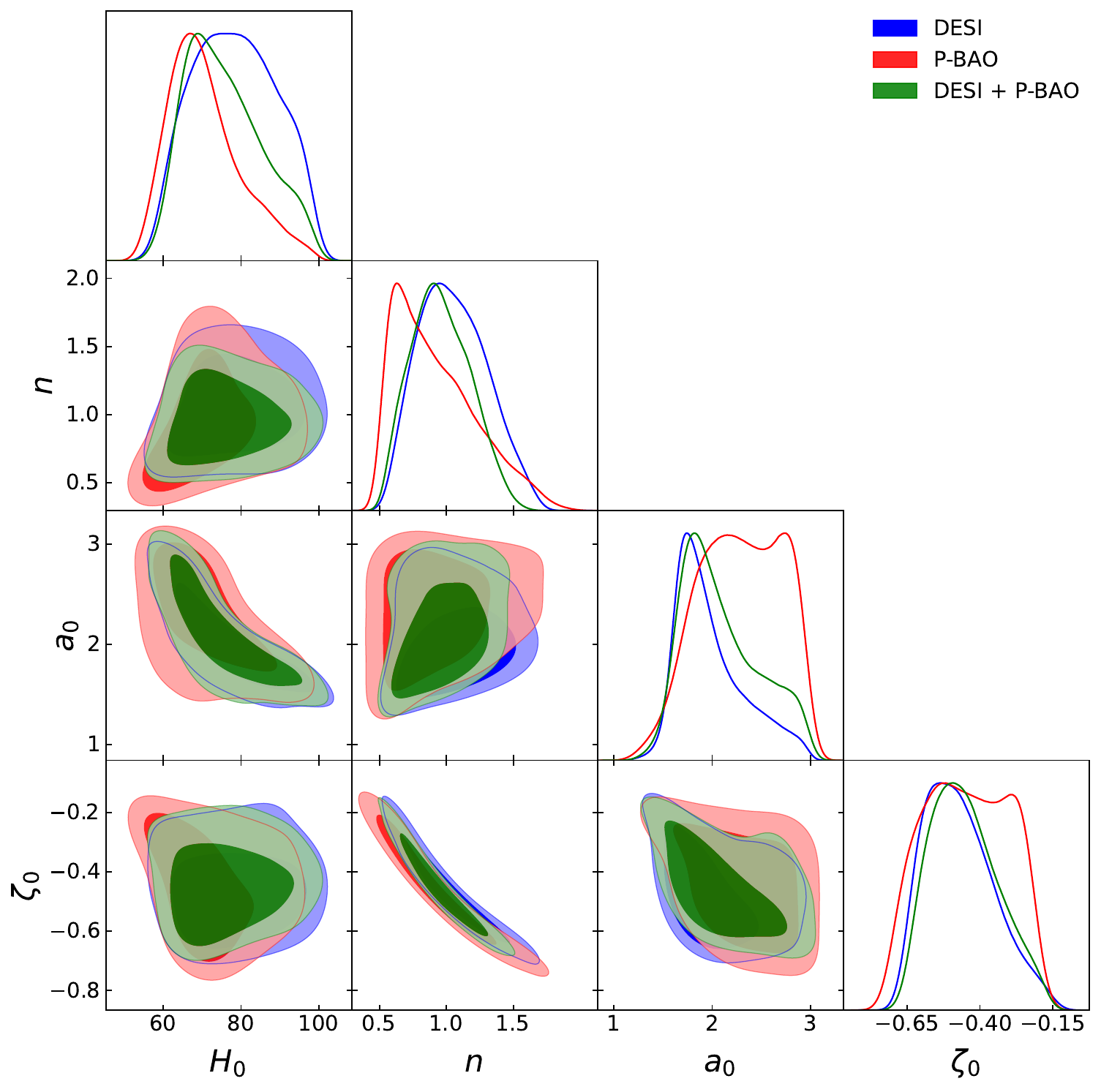}}
\caption{2-d contour sub-plot for the parameters $H_0$, $a_0$, $n$ and $\zeta_0$ with 1-$\sigma$ and 2-$\sigma$ errors (showing the 68\% and 95\% c.l. ) for $D_H(z)$ vs $z$.}
\label{fig12_2}
\end{figure*}

Another parameter that we investigate for our model is the Hubble distance $D_H(z)$. This represents a characteristic distance scale associated with the Hubble expansion rate at redshift $z$. Defined as:
\begin{equation}
D_H(z)=\frac{c}{H(z)},
\end{equation}
where, $c$ is the speed of light in vacuum and $H(z)$ is the Hubble parameter in terms of redshift $z$. Again, we consider two datasets obtained from DESI and previous BAO (P-BAO) observations like SDSS and WiggleZ, as provided in Table \ref{table:7} for the Hubble distance $D_H(z)$. Using which we perform statistical analysis using Markov chain Monte Carlo method for our model, to find the best fitting parameters involved in the Hubble distance $D_H(z)$. To find the mean values of the parameters $H_0$, $a_0$, $n$ and $\zeta_0$ we have used the chi-squared function defined as:
\begin{widetext}
\begin{equation}
\chi^2_{D_H}(H_0,n,a_0,\zeta_0) = \sum_{i=1}^{N} \frac{\left[ {D_H}_{th}(z_i, H_0,n,a_0,\zeta_0) - {D_H}_{obs}(z_i) \right]^2}{\sigma_{D_H}(z_i)^2}.
\label{32b}
\end{equation}
\end{widetext}
Here, $\sigma_{D_H}(z_i)$ denotes the standard error in the observed value of $D_H(z_i)$ at redshift $z_i$, ${D_H}_{th}$ represents the theoretical value of the Hubble distance, and ${D_H}_{obs}$ represents the observed value. By performing MCMC the best fitting model parameters are obtained and results are compared with $\Lambda$CDM. Table \ref{table:8} provides the $R^2$, $\chi^2_{\min}$, AIC and BIC values for the model  under consideration and $\Lambda$CDM for different datasets of $D_H(z)$ vs $z$. The Fig. \ref{fig12} shows effective fit of our model to the observational datasets in comprison to $\Lambda$CDM. Fig. \ref{fig12_2} shows the 2-d contour sub-plot for the parameters $H_0$, $a_0$, $n$ and $\zeta_0$ with 1-$\sigma$ and 2-$\sigma$ errors (showing the 68\% and 95\% c.l.). And, Table \ref{table:8a} shows the best fitting parameter values for $D_H(z)$ vs $z$.\\
Results show $\chi^2_{\min}$, AIC and BIC values for the model in case of each dataset of $D_H(z)$ vs $z$, is sufficiently lower than those of $\Lambda$CDM. Also, in all cases both $\Delta$AIC and $\Delta$BIC values are $>2$ indicating evidence to be favourable to the model under consideration as compared $\Lambda$CDM. The fact that the $R^2$ values in each instance show a somewhat higher goodness of fit with observational data than the $\Lambda$CDM model further demonstrates the overall effectiveness of the model. The results therefore show that the model is able to represent the development of the Hubble distance parameter $D_H(z)$ with redshift $z$ for both the DESI and P-BAO datasets and their combined versions.
\begin{table}[h!]
\centering
\begin{tabular}{|ccc|c||ccc|c|}
\hline
\multicolumn{4}{|c||}{\textbf{DESI}} & \multicolumn{4}{c|}{\textbf{P-BAO}} \\
\hline
$z$ & $D_H/r_d$ & $\sigma_{D_H/r_d}$ & Ref & $z$ & $D_H/r_d$ & $\sigma_{D_H/r_d}$ & Ref \\
\hline
0.698 & 19.77 & 0.47 & \cite{dh56} & 0.510 & 20.98 & 0.61 & \cite{ref98} \\
1.480 & 13.23 & 0.47 & \cite{dh57} & 0.706 & 20.08 & 0.60 & \cite{ref98} \\
2.300 & 9.07  & 0.31 & \cite{dh55} & 0.930 & 17.88 & 0.35 & \cite{ref98} \\
2.400 & 8.94  & 0.22 & \cite{dh60} & 1.317 & 13.82 & 0.42 & \cite{ref98} \\
      &       &      &      & 2.330 & 8.52  & 0.17 & \cite{ref98} \\
\hline
\end{tabular}
\caption{Hubble distance $D_H/r_d$ and their uncertainties at redshift $z$ from the DESI and P-BAO datasets. During analysis The sound horizon at the drag epoch is fixed at $r_d = 147.09 \pm 0.26$ Mpc \cite{dh4}.}
\label{table:7}
\end{table}
\begin{table}[ht]
\centering
\begin{tabular}{|l|c|c|c|}
\hline
\textbf{Statistical Parameters} & \textbf{DESI} & \textbf{P-BAO} & \textbf{DESI+P-BAO} \\
\hline
\multicolumn{4}{|c|}{\textbf{Model}} \\
\hline
$R^2$ Score        & 0.9706 & 0.9980 & 0.9881 \\
$\chi^2_{\min}$    & 7.72   & 1.32   & 9.04  \\
AIC                & 15.72  & 9.32   & 17.04  \\
BIC                & 14.16  & 6.87   & 17.83  \\
\hline
\multicolumn{4}{|c|}{\textbf{$\Lambda$CDM}} \\
\hline
$R^2$ Score        & 0.9575 & 0.9902 & 0.9751 \\
$\chi^2_{\min}$    & 15.63  & 5.93   & 21.56  \\
AIC                & 21.63  & 11.93  & 27.56  \\
BIC                & 20.46  & 10.09  & 28.15  \\
\hline
\multicolumn{4}{|c|}{\textbf{$\Delta$AIC and $\Delta$BIC}} \\
\hline
$\Delta$AIC        & 5.91   & 2.61   & 10.52   \\
$\Delta$BIC        & 6.30   & 3.22   & 10.32   \\
\hline
\end{tabular}
\caption{Statistical comparison between the model and $\Lambda$CDM for different datasets Of $D_H(z)$ vs $z$. $\Delta$AIC and $\Delta$BIC represent the difference ($\Lambda$CDM $-$ Model) for AIC and BIC.}
\label{table:8}
\end{table}
\begin{table*}[ht]
\centering
\begin{tabular}{|l|c|c|c|c|}
\hline
\textbf{Datasets} & $H_0$ & $n$ & $a_0$ & $\zeta_0$ \\
\hline
\textbf{DESI} & $78.31^{+12.27}_{-11.53}$ & $1.032^{+0.290}_{-0.253}$ & $1.891^{+0.469}_{-0.229}$ & $-0.483^{+0.138}_{-0.110}$ \\
\textbf{P-BAO} & $69.30^{+11.13}_{-7.51}$ & $0.854^{+0.398}_{-0.259}$ & $2.306^{+0.471}_{-0.468}$ & $-0.445^{+0.169}_{-0.161}$ \\
\textbf{DESI + P-BAO} & $74.47^{+12.42}_{-8.79}$ & $0.943^{+0.251}_{-0.219}$ & $2.009^{+0.555}_{-0.308}$ & $-0.463^{+0.136}_{-0.110}$ \\
\hline
\end{tabular}
\caption{Best-fit parameter values for the model with 1$\sigma$ uncertainties for different datasets of $D_H(z)$ vs $z$.}
\label{table:8a}
\end{table*} 
\section{Conclusion}\label{sec7}
The primary issue in early universe cosmology has been the intial singularity problem. Subsequent ideas and methods have been developed to overcome this problem. Among them, bouncing cosmology has been put out as a compelling strategy. By altering the curvature or matter components in GR in addition to the EoS, one might get bouncing cosmology. This work discusses the prospect of getting a non-singular bouncing universe scenario in GR that is both physically viable and consistent with the existing cosmological data by introducing a modified EoS. This may be significant as the implementation of a modified EoS might lead to a breach of energy requirements, especially the NEC, which is an essential prerequisite at the bounce point.  And, also latter on can help in expressing the physical viability and consistency with observational data.  In this work, we proposed a novel EoS given as $p=\zeta _0 \rho +\zeta _1 \rho  \left(t-t_0\right){}^{-2 n}$, where $\zeta_0$, $\zeta_1$, $t_0$ and $n$ are some constant parameters.\\
We found the exact solutions to the Friedmann equations in Sec. \ref{sec2} using the proposed form of EoS.  The derived solutions of the model seem to match some of the basic conditions of a scenario of a bouncing universe.  For example, the Hubble parameter disappears at the bouncing epoch, while the scale factor slope is negative during the contraction phase and positive during the expansion phase.  In Sec. \ref{sec3}, we looked at the physical parameters of our model, including the energy density and deceleration parameter.  Both exhibit symmetrical behavior at the bouncing point.  The deceleration parameter goes to negative values after the bounce, and then it enters the phase when decelerated matter dominates and energy density is seen to remain positive through out. We also examined the NEC, SEC, WEC, and DEC evolution in Sec. \ref{sec3} to determine if our model satisfies other necessary requirements, such as violation of energy constraints during the bounce epoch. We found violations of the NEC and SEC at the bouncing point, which are necessary for a non-singular bouncing universe. Additionally, it turns out that the DEC is fulfilled during the bounce. The stability analysis in Sec. \ref{sec4.1} shows that the perturbation terms approach to zero as the time evolves, indicating the model is stable under scalar perturbation. In Sec. \ref{sec4} we investigated the Hubble flow dynamics for the model to examine the model's ability to explain the fundamentals of the inflationary era or the transitions between the matter-bounce and inflation eras. Results showed subsequent values of the model parameters can mediate the model to have the necessary requirement for the inflationary period, along with an exit from the inflation era. From the statefinder  diagnostics in Sec. \ref{sec5}, we show that for various parameter values, the dark energy behaviour of our considered model is Chaplygin Gas-type. Finally, we compare the model with observational datasets in Sec. \ref{sec6}. In comparison with datasets from DESI and earlier BAO (P-BAO) observations like SDSS and WiggleZ, as well as the Pantheon+SH0ES, we examine our model's ability to explain the observational data related to the late-time universe, taking into account analysis of the Hubble parameter, distance modulus, and Hubble distance evolution with redshift. Results demonstrated a somewhat greater goodness of fit for the model with observational data than the $\Lambda$CDM model. Also, statical parameters like AIC and BIC shows evidence favouring the model more. Even with the recently released DESI datasets, the efficacy of fit is noteworthy.\\
We found that the solutions obtained in our work essentially avoid the early Universe's initial singularity by representing a non-singular bouncing scenario. As supplemental features, our model also shows late-time accelerated expansion and inflationary dynamics. Hence, such inhomogeneous time dependent EoS can have potential to drive brief exotic phases around $t_0$ and then relaxes to a constant EoS like fluid, smoothly modeling transient cosmological events without patching separate eras. Even though our model satisfies the conditions for bouncing cosmology, the tensor to scalar ratio, scalar spectral index, and other significant observable characteristics are not included in this paper since they are outside the purview of this investigation. In further study, we would want to investigate this option. For future perspectives, it might also be interesting to analyse how new modified gravity theories, as well as the effects of viscus and barotropic fluids, can be included into the framework.
%% For this sample we use BibTeX plus aasjournalv7.bst to generate the

\end{document}